\documentclass[preprintnumbers,amsmath,amssymb,10pt]{revtex4}

\usepackage{graphicx}
\usepackage{color}
\usepackage{float}
\usepackage{amsmath}
\usepackage{appendix}
\usepackage{subfigure}

\begin{document}

\title{{Theory of two-component superfluidity of microcavity polaritons}}
\author{A.\hspace{0.09cm}Nafis  Arafat$^{1,2}$, Oleg L. Berman$^{1,3}$, Godfrey Gumbs$^{1,2,4}$, and Peter B. Littlewood$^{5,6}$}

\affiliation{$^{1}$The Graduate School and University Center, The
City University of New York, \\
365 Fifth Avenue,  New York, NY 10016, USA\\
$^{2}$Department of Physics and Astronomy, Hunter College of The City University of New York, 695 Park Avenue, \\
$^{3}$Physics Department, New York City College
of Technology,          The City University of New York, \\
300 Jay Street,  Brooklyn, NY 11201, USA \\
New York, NY 10065, USA\\
$^{4}$Donostia International Physics Center (DIPC), P de Manuel Lardizabal, 4, 20018 San Sebastian, Basque Country, Spain \\
$^{5}$ James Franck Institute and Department of Physics,
The University of Chicago, Chicago, Illinois 60637, United States \\
$^{6}$ School of Physics and Astronomy, University of St Andrews, N Haugh, St Andrews KY16 9SS, United Kingdom}

\date{\today}

\begin{abstract}
\noindent
We develop a microscopic mean-field theory describing the coexistence of Bose–Einstein condensates of upper and lower polaritons (UP/LP) in a semiconductor microcavity. Incorporating interbranch scattering within a modified polariton Hamiltonian, we introduce a phenomenological population–split parameter $\alpha$ that quantifies the relative LP/UP occupations. At zero detuning, the critical temperature becomes independent of $\alpha$, converging to a single value that marks the balanced, resonant regime. Away from resonance, variations in $\alpha$ lead to distinctive and experimentally resolvable changes in both the sound velocity $c_s$ and critical temperature $T_c$, relative to the single-component (LP-only) condensate limit. The system under study consists of excitons confined in A transition metal dichalcogenide (TMDC) monolayer, particularly WSe$_2$ embedded within a planar optical microcavity of GaAs where they strongly couple to cavity photons. 
Our analysis focuses on monolayer WSe$_2$ embdedded in a GaAs microcavity. We present results for GaAs/AlGaAs quantum wells embedded in a GaAs microcavity in the Appendix.While mean-field in scope, the framework provides analytic benchmarks and physical insight for future treatments that include dissipation and fluctuations in nonequilibrium polariton superfluids. 
\end{abstract}

\maketitle

\section{Introduction}
\label{sec1}

\medskip
\par
\noindent
Bose–Einstein condensation (BEC) and superfluidity of exciton-polaritons have become key topics in two-dimensional (2D) systems due to the hybrid light–matter nature of polaritons \cite{deng2010,snokeKeeling2017, kavokin2017}, formed via strong coupling between quantum well excitons and cavity photons~\cite{Moskalenko2000,Littlewood2007}. 
Excitons are bound electron–hole pairs that form in semiconductors through Coulomb attraction. When such excitons couple strongly to the quantized electromagnetic field of an optical microcavity, they hybridize with cavity photons to form new light–matter quasiparticles known as exciton–polaritons \cite{weisbuch1992}. In GaAs/AlGaAs structures, the excitons are confined within a quantum well embedded between distributed Bragg reflectors that define the optical cavity \cite{Skol1998}. In transition metal dichalcogenide (TMDC) realizations, by contrast, the excitons reside in an atomically thin monolayer placed inside a similar cavity, where the large binding energy enhances the strength of light–matter coupling \cite{dufferwiel2015}. Their low effective mass enables high critical temperatures for superfluidity ~\cite{Carusotto2013,Proukakis2017,Peng2022}, with robust behavior under disorder and external fields~\cite{berman2006}. Polaritons support parametric scattering~\cite{ciuti2003, savona2005} 
and coherence across diverse materials, including GaAs~\cite{berman2008}, 
CdTe~\cite{richard2005}, and TMDCs~\cite{manzeli2017,lerario2017}, where 
Bose-Einstein condensation~\cite{kasprzak2006} and room-temperature superfluidity have been observed~\cite{dufferwiel2015,kogar2017,Peng2022,
zhao2021}, alongside promising optoelectronic applications \cite{mak2010} and the emergence of topological excitations in spinor condensates~\cite{rubo2007}. Theoretical models have further explored nonlinear polariton condensation beyond the linear regime~\cite{eastham2001}. Additionally, experiments~\cite{Comaron2025} have highlighted how tunable polariton interactions in driven-dissipative systems enhance spatial coherence and enable Berezinskii--Kosterlitz--Thouless-like crossovers, consistent with interaction-driven superfluid behavior.

\medskip
\par
\noindent
The study of polariton condensates naturally extends to a broader class of multi-component quantum fluids that exhibit rich collective phenomena. Multi-component quantum condensates (comprising coexisting, interacting macroscopic quantum states) exhibit collective phenomena absent in single-component superfluids. In spinor atomic Bose–Einstein condensates, simultaneous occupation of multiple hyperfine states gives rise to coupled spin-density modes, relative-phase dynamics, and topological defects such as half-quantum vortices~\cite{Ho1998, Stamper2013}. In magnon systems, multi-branch condensation produces hybridized magnon modes and interaction-mediated phase transitions~\cite{Demokritov2006, Aeppli2022}.An instructive parallel can be drawn with superfluid $^3$He, where distinct pairing symmetries (A and B phases) correspond to multiple order parameters with coupled collective excitations~\cite{Vollhardt2003, Leggett1975}. Although the microscopic mechanisms differ, the analogy underscores how coexisting condensate components can support richer excitation spectra and phase dynamics than their single-component counterparts.
\medskip
\par
\noindent
Microcavity polaritons offer a uniquely tunable platform for exploring two-component condensate physics. The upper (UP) and lower (LP) polariton branches arise naturally from strong exciton–photon coupling, with an energy splitting that can be engineered through cavity design. 
These two branches correspond to the eigenmodes of the coupled exciton–photon system obtained through the Hopfield diagonalization. 
The lower polariton (LP) has a larger excitonic fraction near positive detuning, giving it a heavier effective mass and stronger interactions, whereas the upper polariton (UP) becomes more photonic at negative detuning and has a lighter mass. 
The two modes are separated by the vacuum Rabi splitting $\hbar\Omega_R$, which quantifies the strength of the light–matter coupling and sets the energy scale for the polariton system~\cite{Carusotto2013,weisbuch1992}. While most experiments observe condensation exclusively in the lower branch due to efficient phonon-assisted relaxation \cite{kasprzak2006}, recent experimental work has demonstrated upper-branch condensation~\cite{Chen2023}, suggesting that controlled dual-branch occupation may be achievable under appropriate conditions. A key control parameter in such systems is the exciton–photon detuning, $\Delta_0$, defined as the energy difference  between the bare cavity photon mode and the exciton resonance. Detuning can be tuned in situ by adjusting the cavity length, refractive index (via temperature or angle of incidence), or through fabrication of distributed Bragg reflectors of different thickness \cite{dufferwiel2015}. Varying $\Delta_0$ changes the Hopfield coefficients that determine the excitonic and photonic fractions of the polariton branches, thereby modifying their effective masses and interaction strengths. Near zero detuning, both components are equally mixed, while large positive or negative detuning yields predominantly excitonic or photonic character, respectively. This tunability, along with control over excitation density and bare Rabi splitting, allows polaritons to access regimes ranging from light-mass, weakly interacting fluids to heavy-mass, strongly interacting excitonic gases.
\medskip
\par
\noindent
Despite these experimental advances and the theoretical progress in understanding single-branch polariton condensates, no microscopic theory has yet incorporated mutual coherence and interaction between lower and upper polariton (LP and UP) condensates in a unified superfluid framework. The upper branch, despite experimental signatures of occupation~\cite{Chen2023}, remains largely absent from such treatments. Previous work has focused largely on one-component polariton condensates, particularly in the lower polariton (LP) branch. Polariton condensates can form in two distinct branches---lower polariton (LP) and upper polariton (UP)---arising from Rabi splitting~\cite{szymanska2004,snokecoherent2006}. 

\medskip
\par
\noindent
Motivated by advances in multi-component superfluidity and tunable quasiparticle engineering~\cite{arafat2024semi,Sanvitto2016,alyatkin2020,Bhattacharya2013}, we develop a theory of superfluidity in a two-component polariton condensate comprising both LP and UP branches, formulated within an equilibrium mean-field (Bogoliubov) framework that explicitly includes detuning dependence and interaction-driven effects. Our analysis is carried out in the long-wavelength limit, where the lower and upper polariton dispersions are expanded to quadratic order in momentum, and the effective masses treated as constants, allowing analytical access to the collective modes and superfluid parameters such as the concentration of the superfluid component and the critical temperature for superfluidity. Our theoretical framework treats the condensate population distribution between LP and UP branches as a phenomenological input, characterized by a parameter $\alpha \in [0,1]$, and calculates the resulting superfluid properties via mean-field theory. The systems we consider in particular are GaAs/AlGaAs quantum wells embedded in a GaAs microcavity, and TMDC (WSe$_2$) monolayers embedded in a GaAs microcavity~\cite{regan2022}. Our work establishes theoretical benchmarks for an experimentally relevant regime, soon accessible with current advances in polariton microcavities. We provide quantitative predictions for the collective excitation spectrum, sound velocity, and critical temperature for superfluidity that would uniquely distinguish genuine two-component LP–UP superfluidity from single-branch superfluidity, thereby offering clear signatures for future experimental verification.

\medskip
\par
\noindent
In this main text we present a fair amount of results for monolayer transition metal dichalcogenides (TMDCs), 
which exhibit exceptional excitonic binding energies and ultralight 
effective masses. Monolayers such as MoS$_2$, MoSe$_2$, WS$_2$, and WSe$_2$ are 
direct-gap semiconductors~\cite{splendiani2010,mak2010} with strong 
light–matter coupling~\cite{liu2015}. Their hexagonal Brillouin zones 
host valley-selective optical transitions at the K and K$^\prime$ points~\cite{
xiao2012,xu2014}, described by k$\cdot$p theory~\cite{Kormanyos2015}, and 
their excitons possess large binding energies and long radiative 
lifetimes~\cite{wang2018}. Exciton–polariton condensation and even 
room-temperature superfluidity have been reported in TMDC microcavities~
\cite{kasprzak2006,dufferwiel2015,kogar2017,Peng2022,zhao2021}, while 
doping enables charged trions~\cite{Mak2013}, and engineered 
heterostructures support interlayer excitons~\cite{rivera2015,hong2014} 
and dipolar and time-crystalline phases phases~\cite{berman2022nanomaterials,martins2023}. Related excitonic phenomena have also emerged in topological 
semimetals~\cite{kandel2023,kandel2024}, underscoring the universality 
of strongly coupled light–matter fluids across material platforms.

\medskip
\par
\noindent
We show that this two-component structure yields enhanced and tunable sound velocity and critical temperature of superfluidity compared to the LP-only case.  Our results are general, material-independent, and analytically tractable—offering a predictive framework for future studies of nonequilibrium polariton dynamics, Lindblad-driven condensates, and Josephson-like phase control in strongly coupled light–matter systems.

\medskip
\par
\noindent
The remainder of this paper is organized as follows.
Section~II introduces the modified Hamiltonian for upper and lower polaritons, including the systematic reduction of interaction terms.
Section~III derives the collective excitation spectrum within the Bogoliubov framework. Section~IV examines the tunable sound velocity and its dependence on detuning, Rabi splitting, and the population–split parameter $\alpha$. Section~V develops the superfluid criterion and derives analytic expressions for the critical temperature for superfluidity and concentration. In this section we also analyze how the critical temperature for superfluidity depends on the detuning, Rabi splitting, and population-split parameter $\alpha$, and demonstrate a retrieval procedure that allows one to infer $\alpha$ from measured values of the critical temperature $T_c$ and critical density $n_c$. Section~VI concludes with a discussion of the main findings and their implications for tunable two-component polariton superfluidity.  Appendices~A and~B present detailed derivations of the interaction reduction and polariton effective masses that are used in our calculations. Appendix C presents all pertinent results of our analysis for GaAs/AlGaAs embedded in GaAs microcavities.

\section{Effective Hamiltonian of upper and lower polaritons}
\label{sec2}

\medskip
\par
\noindent
In this section, we modify the Hamiltonian for our system of excitons and photons to account for both upper and lower polaritons. By analyzing this modified Hamiltonian, we aim to study the critical temperature for Bose-Einstein condensation and the superfluidity of a two-component polariton system, focusing on the role played by scattering between the polariton branches and the collective excitations in the system.

\medskip
\par
\noindent
Polaritons are quantum superpositions of excitons and photons. The total Hamiltonian of excitons and microcavity photons is given by $\hat{H}_{\text{total}} ~=~ \hat{H}_{\text{exc}} + \hat{H}_{\text{ph}} + \hat{H}_{\text{exc-ph}}$, where $\hat{H}_{\text{exc}}$ is the excitonic Hamiltonian, $\hat{H}_{\text{ph}}$ is the photonic Hamiltonian, and $\hat{H}_{\text{exc-ph}}$ is a Hamiltonian of exciton-photon interaction. We can define each of these explicitly.

\medskip
\par
\noindent
We begin with the Hamiltonian for 2D excitons in an infinite homogeneous system, given by

\begin{equation}
\hat{H}_{\text{exc}} = \sum_P \varepsilon_{\text{ex}}(P) \hat{b}_P^\dagger \hat{b}_P + \frac{1}{2A} \sum_{\substack{P,P',q}} U_q \hat{b}_{P+q}^\dagger \hat{b}_{P'-q}^\dagger \hat{b}_P \hat{b}_{P'} \  ,
\end{equation}

\medskip
\par
\noindent
where $A$ is a normalization area. 
Here, the first term represents the energy of the excitons, and the second term represents the exciton-exciton interaction. The operator \( \hat{b}_P \) is the excitonic annihilation operator, and the operator \( \hat{b}^{\dagger}_P \) is the excitonic creation operator for momentum \( P \). The contact potential $U_q$ is approximated by Eq. (\ref{InteractionContact}) and from  experimental benchmarks. The exciton energy dispersion $\varepsilon_{\text{ex}}(P)$ is given by

\begin{equation}
\varepsilon_{\text{ex}}(P) = E_{\text{band}} - E_{\text{binding}} + \varepsilon_0(P) \ ,
\label{PhotonDisp}
\end{equation}
The Hamiltonian for noninteracting microcavity photons is given by 

\begin{equation}
\hat{H}_{\text{ph}} = \sum_P \varepsilon_{\text{ph}}(P) \hat{a}_P^\dagger \hat{a}_P \ \ .
\label{PhotonHamiltonian}
\end{equation}
Here, the operator \( \hat{a}_P \) is the photonic annihilation operator, and the operator \( \hat{a}^{\dagger}_P \) is the photonic creation operator for momentum \( P \). For photons confined to a microcavity, the energy dispersion $\varepsilon_{\text{ph}}(P) $ is given by

\begin{equation}
\varepsilon_{\text{ph}}(P) = \frac{c}{n} \sqrt{P^2 + \hbar^2 \pi^2 L_c^{-2}} \ .
\end{equation}

\noindent
where $L_c$ is the microcavity length. The exciton-photon interaction term is given by

\begin{equation}
\hat{H}_{\text{exc-ph}} = \sum_P \hbar \Omega_R (\hat{b}_P^\dagger \hat{a}_P + \hat{a}_P^\dagger \hat{b}_P),
\end{equation}

\medskip
\par
\noindent
where \( \hbar \Omega_R \) is the Rabi splitting, representing the coupling strength between photons and excitons.To fully describe the system, we express the Hamiltonian in the basis of upper and lower polariton operators. We substitute the following expressions from Eq. (6) of Ref. \cite{hopfield1958}. 

\begin{equation}
\hat{b}_P = X_P \hat{l}_P - C_P \hat{u}_P, \quad \hat{b}_P^\dagger = X_P \hat{l}_P^\dagger - C_P \hat{u}_P^\dagger
\end{equation}

\noindent
where \( \hat{l}_P \) is the lower polariton annihilation operator for momentum \( P \), \( \hat{u}_P \) is the upper polariton annihilation operator for momentum \( P \), \( X_P \) is the excitonic fraction of the lower polariton and \( C_P \) is the photonic fraction of the lower polariton. The photonic operators may be defined in the basis of upper and lower polariton operators as 

\begin{equation}
\hat{a}_P = C_P \hat{l}_P + X_P \hat{u}_P, \quad \hat{a}_P^\dagger = C_P \hat{l}_P^\dagger + X_P \hat{u}_P^\dagger.
\label{photon_operators}
\end{equation}

\noindent
The photon and exciton fractions in the lower polariton branch, denoted as \( X_P \) and \( C_P \) respectively, describe the relative contributions of the photonic and excitonic components in the polariton wavefunction. These fractions are given by

\begin{equation}
X_P = \frac{1}{\sqrt{1 + \left( \frac{\hbar \Omega_R}{\varepsilon_{\text{LP}}(P) - \varepsilon_{\text{ph}}(P)} \right)^2}},
\end{equation}

\begin{equation}
C_P = -\frac{\frac{\hbar \Omega_R}{\varepsilon_{\text{LP}}(P) - \varepsilon_{\text{ph}}(P)}}{\sqrt{1 + \left( \frac{\hbar \Omega_R}{\varepsilon_{\text{LP}}(P) - \varepsilon_{\text{ph}}(P)} \right)^2}},
\end{equation}

\noindent
where \( \varepsilon_{\text{LP}}(P) \) is the energy of the lower polariton branch,
\( \varepsilon_{\text{ph}}(P) \) is the photon dispersion. The polariton energy is given by

\begin{equation}
\varepsilon_{\text{LP/UP}}(P) = \frac{\varepsilon_{\text{ph}}(P) + \varepsilon_{\text{exc}}(P)}{2} \mp \frac{1}{2} \sqrt{\left( \varepsilon_{\text{ph}}(P) - \varepsilon_{\text{exc}}(P) \right)^2 + 4| \hbar \Omega_R|^2} \ ,
\label{UpLowS1}
\end{equation}

\medskip
\par
\noindent
The linear part of the Hamiltonian, which includes the linear part of $\hat{H}_{\text{exc}}$ and $\hat{H}_{\text{ph}}$, and the interaction between exciton and photon $\hat{H}_{\text{exc-ph}}$ can be diagonalized by applying unitary transformations and has the form

\begin{equation}
\hat{H}_0 = \sum_{P} \varepsilon_{LP} (P) \hat{l}_{P}^{\dagger} \hat{l}_{P} + \sum_{P} \varepsilon_{UP} (P) \hat{u}_{P}^{\dagger} \hat{u}_{P}
\label{noninteractingHam}
\end{equation}

\medskip
\par
\noindent
The next thing we need to consider are interactions.
We substitute the expressions for \( \hat{b}_P \) and \( \hat{b}_P^\dagger \) into the excitonic interaction term,

\[
\frac{1}{2A} \sum_{\substack{P,P',q}} U_q \hat{b}_{P+q}^\dagger \hat{b}_{P'-q}^\dagger \hat{b}_P \hat{b}_{P'}
\]

\medskip
\par
\noindent
where, the contact potential $U_q$ for the exciton-exciton interaction is approximated to $U_0$, and is defined for GaAs/AlGaAs as \cite{Carusotto2013}

\begin{equation}
U_0 = \frac{6e^2 a_{2D}}{\epsilon}
\label{InteractionContact}
\end{equation} 

\medskip
\par
\noindent
and $a_{2D}$ is the effective 2D Bohr radius of excitons. 

\begin{equation}
a_{2D} = \frac{\hbar^2 \epsilon}{2\mu_{e-h} e^2}
\end{equation}
\medskip
\par
\noindent

\noindent
It is crucial to note that for monolayer WSe$_2$, the exciton–exciton interaction is more accurately described by the Rytova-Keldysh potential~\cite{Berkelbach2013}, 
which accounts for reduced screening in atomically thin materials. Rather than deriving $U_0$ microscopically, we adopt the experimentally measured value from polariton blueshift measurements~\cite{Kravtsov2020,
flatten2016}. Proceeding we get the following

\begin{equation}
\frac{1}{2A} \sum_{\substack{P,P',q}} U_0 \left( X_{P+q} \hat{l}_{P+q}^\dagger - C_{P+q} \hat{u}_{P+q}^\dagger \right) \left( X_{P'-q} \hat{l}_{P'-q}^\dagger - C_{P'-q} \hat{u}_{P'-q}^\dagger \right) \left( X_P \hat{l}_P - C_P \hat{u}_P \right) \left( X_{P'} \hat{l}_{P'} - C_{P'} \hat{u}_{P'} \right)
\end{equation}
\medskip
\par
\noindent
Expanding this product results in terms involving, pure lower polariton interactions, pure upper polariton interactions, cross-interactions between upper and lower polaritons. We start with the expression

\[
\frac{1}{2A} \sum_{\substack{P,P',q}} U_0 \left( X_{P+q} \hat{l}_{P+q}^\dagger - C_{P+q} \hat{u}_{P+q}^\dagger \right) \left( X_{P'-q} \hat{l}_{P'-q}^\dagger - C_{P'-q} \hat{u}_{P'-q}^\dagger \right) \left( X_P \hat{l}_P - C_P \hat{u}_P \right) \left( X_{P'} \hat{l}_{P'} - C_{P'} \hat{u}_{P'} \right)
\]

\medskip
\par
\noindent
The subsequent 16 interaction terms can be combined into a single equation, in a more reasonable order, grouping together terms that are alike. We explicitly sort these terms in Appendix \ref{InteractionReduction}.

\medskip
\par
\noindent
Certain terms in the expanded Hamiltonian represent interactions that are not typical for polaritons due to the following reasons. Polaritons in the upper (\( \hat{u} \)) and lower (\( \hat{l} \)) branches generally do not transform into one another without a strong coupling mechanism, especially with the large Rabi splitting. The Rabi splitting represents the energy difference between the upper and lower polariton branches, and when this splitting is significant (on the order of tens of meV), transitions between these branches require a comparable amount of energy, making spontaneous interbranch conversions highly improbable~\cite{Peng2022,Carusotto2013}. As a result, terms where a lower and an upper polariton combine to form two of the same type (either both lower or both upper) are not physically consistent and are sorted through in Appendix A. Processes that would convert one lower and one upper polariton into two lower or two upper polaritons are unlikely to occur without the input of additional energy. This makes certain cross-type conversion terms energetically unfavorable, if even possible, so we discard them from consideration. After removing these non-physical terms, the interaction Hamiltonian simplifies to

\begin{equation}
\begin{aligned}
\hat{H}_{\text{int}} = \frac{1}{2A} \sum_{\substack{P, P', q}} U_0 \Big[ 
& g_{\ell\ell}^{PP'q} \,
\hat{l}_{P+q}^\dagger \hat{l}_{P'-q}^\dagger \hat{l}_{P} \hat{l}_{P'} 
+ g_{uu}^{PP'q} \,
\hat{u}_{P+q}^\dagger \hat{u}_{P'-q}^\dagger \hat{u}_{P} \hat{u}_{P'} \\
& + g_{\ell u}^{PP'q} \,
\hat{l}_{P+q}^\dagger \hat{u}_{P'-q}^\dagger \hat{l}_{P} \hat{u}_{P'}
+ g_{u\ell}^{PP'q} \,
\hat{u}_{P+q}^\dagger \hat{l}_{P'-q}^\dagger \hat{u}_{P} \hat{l}_{P'} \\
& + \tilde g_{\ell u}^{PP'q} \,
\hat{l}_{P+q}^\dagger \hat{u}_{P'-q}^\dagger \hat{u}_{P} \hat{l}_{P'}
+ \tilde g_{u\ell}^{PP'q} \,
\hat{u}_{P+q}^\dagger \hat{l}_{P'-q}^\dagger \hat{l}_{P} \hat{u}_{P'} 
\Big].
\label{interactionH}
\end{aligned}
\end{equation}

\noindent
With the Hopfield factor coefficients defined as

\begin{align}
g_{\ell\ell}^{PP'q} &= X_{P+q} X_{P'-q} X_{P} X_{P'}, \\
g_{uu}^{PP'q} &= C_{P+q} C_{P'-q} C_{P} C_{P'}, \\
g_{\ell u}^{PP'q} &= X_{P+q} C_{P'-q} X_{P} C_{P'}, \\
g_{u\ell}^{PP'q} &= C_{P+q} X_{P'-q} C_{P} X_{P'}, \\
\tilde g_{\ell u}^{PP'q} &= X_{P+q} C_{P'-q} C_{P} X_{P'}, \\
\tilde g_{u\ell}^{PP'q} &= C_{P+q} X_{P'-q} X_{P} C_{P'}.
\end{align}

\medskip
\par
\noindent
To reiterate, the total Hamiltonian in the upper and lower polariton basis then becomes

\begin{equation}
\begin{aligned}
\hat{H}_{\text{total}}
&= \hat{H}_{0} + \hat{H}_{\text{int}} \\[6pt]
&= \sum_{P} 
    \varepsilon_{\text{LP}}(P) \,
    \hat{l}_{P}^{\dagger} \hat{l}_{P}
 \;+\; \sum_{P} 
    \varepsilon_{\text{UP}}(P) \,
    \hat{u}_{P}^{\dagger} \hat{u}_{P} \\[6pt]
&\quad + \frac{1}{2A} 
   \sum_{\substack{P, P', q}} 
   U_q \Bigg\{ \,
      g_{\ell\ell}^{PP'q} \,
      \hat{l}_{P+q}^\dagger 
      \hat{l}_{P'-q}^\dagger 
      \hat{l}_{P} \hat{l}_{P'} \\[3pt]
&\qquad\qquad\qquad
    +\, g_{uu}^{PP'q} \,
      \hat{u}_{P+q}^\dagger 
      \hat{u}_{P'-q}^\dagger 
      \hat{u}_{P} \hat{u}_{P'} \\[3pt]
&\qquad\qquad\qquad
    +\, g_{\ell u}^{PP'q} \,
      \hat{l}_{P+q}^\dagger 
      \hat{u}_{P'-q}^\dagger 
      \hat{l}_{P} \hat{u}_{P'} \\[3pt]
&\qquad\qquad\qquad
    +\, g_{u\ell}^{PP'q} \,
      \hat{u}_{P+q}^\dagger 
      \hat{l}_{P'-q}^\dagger 
      \hat{u}_{P} \hat{l}_{P'} \\[3pt]
&\qquad\qquad\qquad
    +\, \tilde g_{\ell u}^{PP'q} \,
      \hat{l}_{P+q}^\dagger 
      \hat{u}_{P'-q}^\dagger 
      \hat{u}_{P} \hat{l}_{P'} \\[3pt]
&\qquad\qquad\qquad
    +\, \tilde g_{u\ell}^{PP'q} \,
      \hat{u}_{P+q}^\dagger 
      \hat{l}_{P'-q}^\dagger 
      \hat{l}_{P} \hat{u}_{P'}
   \Bigg\}.
\label{FinalHamiltonianKspace}
\end{aligned}
\end{equation}

\noindent
We emphasize that the interaction coefficients $g_{\ell\ell}, g_{uu}, g_{\ell u}, g_{u\ell}, \tilde g_{\ell u}, \tilde g_{u\ell}$ defined above are in general momentum-dependent through the Hopfield fractions $X_{P}$ and $C_{P}$. Retaining the full momentum dependence ensures that inter- and intra-branch scattering are treated rigorously across all detunings and Rabi couplings. However, such generality comes at the cost of analytic tractability: closed-form expressions for the Bogoliubov spectrum, sound velocity, and critical temperature become intractable if one carries the momentum dependence throughout. For this reason, we adopt a controlled approximation appropriate for condensates forming near $P = 0$. We thus evaluate the Hopfield coefficients at zero in-plane momentum, $X_P \to X_0$ and $C_P \to C_0$, and use these values to define effective interaction strengths. This approximation is physically justified because Bose-Einstein condensation occurs predominantly at zero momentum, where the single-particle energies are minimized and superfluid properties such as sound velocity and critical temperature are governed by low-energy, long-wavelength collective modes rather than high-momentum excitations. Additionally, at small $P$, the Hopfield coefficients vary slowly, so $X_P \approx X_0$ and $C_P \approx C_0$ remain accurate.

\section{Spectrum of collective excitations}

In analyzing the excitation spectrum of the coexisting LP-UP condensate, we may begin in the long-wavelength limit ($P \to 0$) where the dispersions can be linearized and the effective polariton masses treated as constants. This is the regime directly relevant for superfluid properties such as the sound velocity and the critical temperature. Both quantities are governed by the low-energy, phonon-like Bogoliubov modes rather than by high-momentum single-particle excitations. Moreover, at long wavelengths the momentum dependence of the Hopfield coefficients becomes negligible, allowing the interaction Hamiltonian to be expressed in terms of constant effective couplings \cite{AGD2012}. This approximation preserves the essential collective physics while avoiding artifacts introduced by momentum-dependent mass renormalization, and is standard practice when applying the Landau criterion or deriving analytic estimates for $c_s$ and $T_c$.

\medskip
\par
\noindent
We define \( \Delta_0 = \frac{c}{n} \hbar \pi L_c^{-1} - \left( E_{\text{band}} - E_{\text{binding}} \right) \) as the detuning. At $P=0$ the Hopfield coefficients become momentum independent,
\begin{equation}
X_{P}\to X_0,\qquad C_{P}\to C_0,
\quad
X_0^2=\tfrac12\!\left(1+\frac{\Delta_0}{\sqrt{\Delta_0^2+4(\hbar\Omega_R)^2}}\right),\ 
C_0^2=1-X_0^2,
\label{HopfieldEqns}
\end{equation}

Crucially, the Hopfield coefficients retain their full detuning and Rabi-splitting dependence. This allows us to explore off-resonance regimes while maintaining analytic tractability. The resulting expressions for collective modes, sound velocity, and critical temperature are exact within the long-wavelength limit and provide physically faithful benchmarks for comparison with experiments and future numerical treatments that incorporate full momentum dependence. The Hopfield coefficients at  $P=0$ can be varied widely by varying the detuning and the Rabi splitting as seen below in Figure~\ref{HopfieldGraphs}.

\begin{figure}[H]
    \centering
    \includegraphics[width=0.6\linewidth]{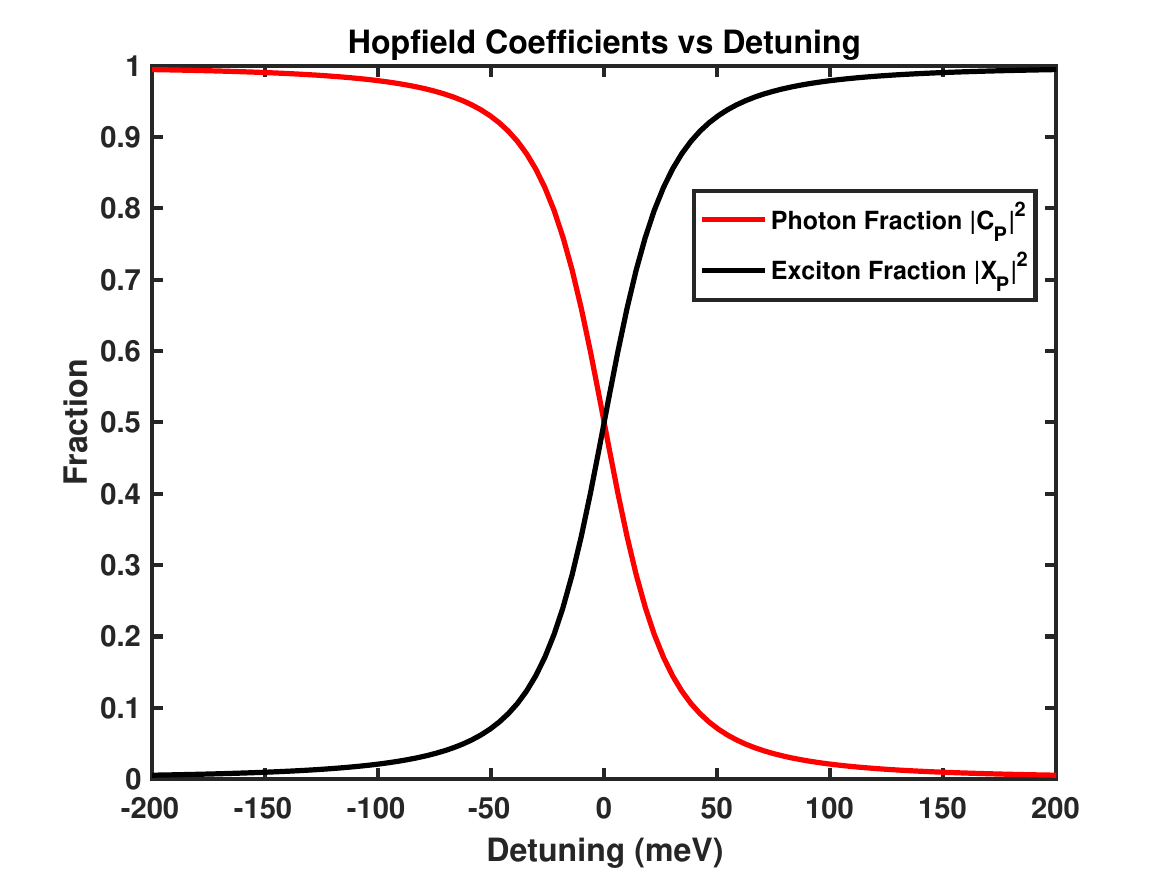}
    \caption{The Hopfield coefficients as a function of the detuning $\Delta_0$ defined above for WSe$_2$. We employed the definitions in Eq.~(\ref{HopfieldEqns}). We kept the Rabi splitting constant at $\Omega_R = $ 8 meV.}
    \label{HopfieldGraphs}
\end{figure}

\noindent
For the contact interaction we take $U_{q}\to U_0$ using Eq.~(\ref{InteractionContact}) for GaAs/AlGaAs quantum wells, and the experimentally measured value from polariton blueshift measurements for monolayers of WSe$_2$ ~\cite{Kravtsov2020,
flatten2016}. We then define the constant effective couplings at $P = 0$
\begin{equation}
g_{\ell\ell}\equiv U_0 |X_0|^4,\qquad
g_{uu}\equiv U_0  |C_0|^4,\qquad
g_{\ell u}\equiv U_0 |X_0|^2 |C_0|^2,
\label{eq:const_couplings_P0}
\end{equation}
the interaction Hamiltonian in the polariton basis reduces to
\begin{equation}
\begin{aligned}
\hat{H}_{\text{int}}^{(P=0)} 
= \frac{1}{2A}\sum_{\substack{P,P',q}}
\Big[
& g_{\ell\ell}\,
\hat{l}_{P+q}^\dagger \hat{l}_{P'-q}^\dagger \hat{l}_{P} \hat{l}_{P'}
+ g_{uu}\,
\hat{u}_{P+q}^\dagger \hat{u}_{P'-q}^\dagger \hat{u}_{P} \hat{u}_{P'} \\
& + g_{\ell u}\,
\hat{l}_{P+q}^\dagger \hat{u}_{P'-q}^\dagger \hat{l}_{P} \hat{u}_{P'}
+ g_{\ell u}\,
\hat{u}_{P+q}^\dagger \hat{l}_{P'-q}^\dagger \hat{u}_{P} \hat{l}_{P'} \\
& + g_{\ell u}\,
\hat{l}_{P+q}^\dagger \hat{u}_{P'-q}^\dagger \hat{u}_{P} \hat{l}_{P'}
+ g_{\ell u}\,
\hat{u}_{P+q}^\dagger \hat{l}_{P'-q}^\dagger \hat{l}_{P} \hat{u}_{P'}
\Big].
\label{eq:Hint_P0}
\end{aligned}
\end{equation}
\noindent
The non-interacting contribution to the total Hamiltonian is given by Eq.~(\ref{noninteractingHam}), where the single-particle dispersions $\varepsilon_{\mathrm{LP/UP}}$ are defined in Appendix B, Eq.~(\ref{UPLPLin}). Collecting all terms, we obtain

\begin{equation}
\boxed{
\hat{H}_{\text{total}}^{(P=0)}=\hat{H}_0+\hat{H}_{\text{int}}^{(P=0)}
},
\label{eq:Htotal_P0}
\end{equation}
we obtain the long-wavelength effective Hamiltonian we will use to derive the collective excitation spectrum.

\medskip
\par
\noindent
At $T = 0$ K we assume that both lower and upper polaritons macroscopically occupy the zero-momentum state, forming a coexisting condensate. Expanding the total Hamiltonian in fluctuations around the condensates and applying the Bogoliubov approximation, we obtain the quadratic Hamiltonian governing the collective excitations. Diagonalization yields two excitation branches with dispersion
\begin{equation}
\Omega_{j,P} = \sqrt{\frac{\omega_{\text{LP},P}^2 + \omega_{\text{UP},P}^2 + (-1)^{j-1}\sqrt{\left(\omega_{\text{LP},P}^2 - \omega_{\text{UP},P}^2\right)^2 + \left(4 G_{\text{LP-UP}}\right)^2 \varepsilon_{\text{LP}}(P) \varepsilon_{\text{UP}}(P)}}{2}} \ ,
\label{eq:collective_spectrum_hopfield}
\end{equation}
where $j = 1,2$ and
\begin{align}
\omega_{\text{LP},P}^2 &= \varepsilon_{\text{LP}}(P)^2 + 2 G_{\text{LP}} \, \varepsilon_{\text{LP}}(P), \label{eq:omega_lp}\\
\omega_{\text{UP},P}^2 &= \varepsilon_{\text{UP}}(P)^2 + 2 G_{\text{UP}} \, \varepsilon_{\text{UP}}(P), \label{eq:omega_up}
\end{align}
with interaction parameters

\begin{align}
G_{\text{LP}} &= g_{\ell\ell}\, n_{\text{LP},0}, \quad g_{\ell\ell} = U_0 |X_0|^4, \\
G_{\text{UP}} &= g_{uu}\, n_{\text{UP},0}, \quad g_{uu} = U_0 |C_0|^4, \\
G_{\text{LP-UP}} &= g_{\ell u}\, \sqrt{n_{\text{LP},0} n_{\text{UP},0}}, \quad g_{\ell u} = U_0 |X_0|^2 |C_0|^2.
\label{IntParameters}
\end{align}

\begin{figure}[H]
    \centering    \includegraphics[width=0.65\linewidth]{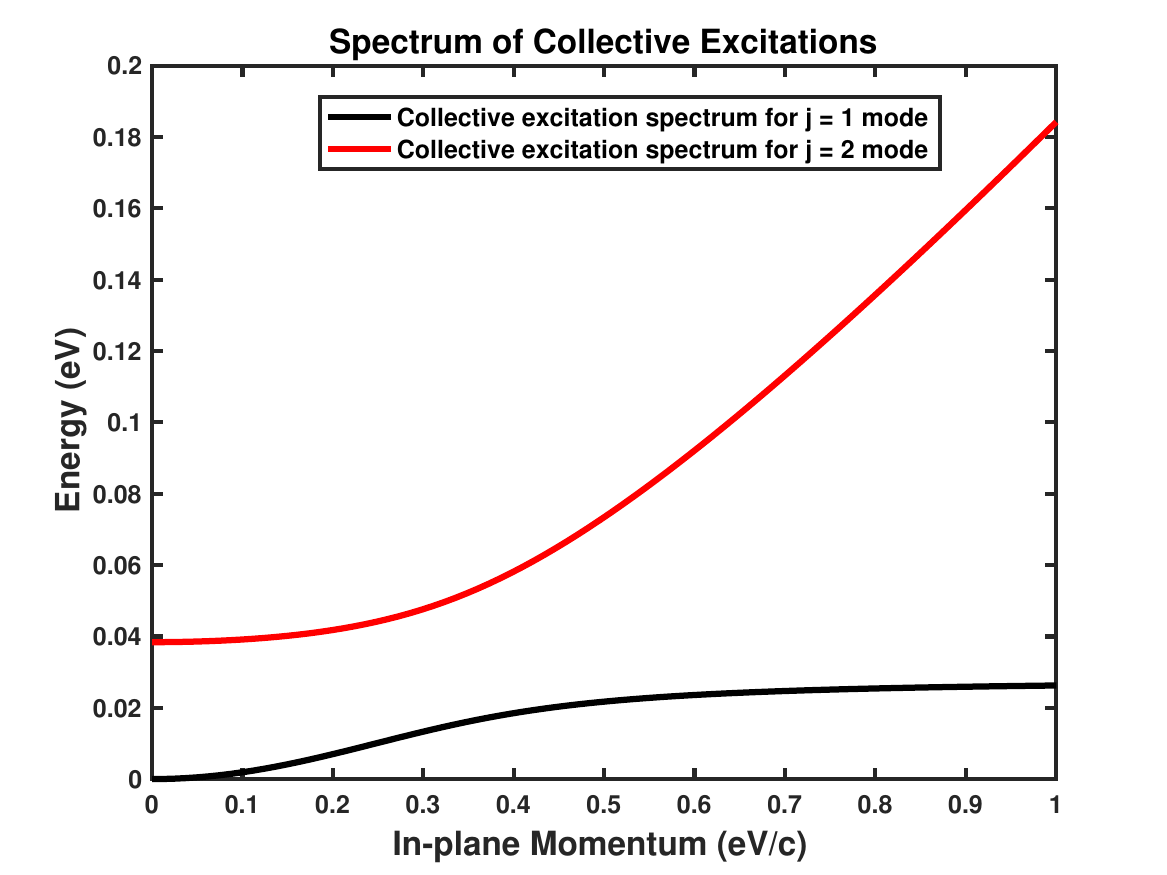}
\caption{The spectrum of collective excitations  in GaAs as a function of momentum. The parameters used in the calculation are $\Omega_R = 17 $ meV, a bare photon of energy 1.6~eV, and a detuning  $\Delta_0 = 0$ meV. The collective excitation spectrum is defined in Eq.\  (\ref{eq:collective_spectrum_hopfield}).}
    \label{CollectiveExcitations}
\end{figure}

\noindent
The two-mode Bogoliubov spectrum for coupled condensates has been studied extensively in atomic systems~\cite{Pindzola}, where hybridization between branches arises from interspecies interactions. Here $n_{\mathrm{LP},0}$ and $n_{\mathrm{UP},0}$ denote the polariton densities in the lower and upper polariton branches, $U_0$ is the contact exciton-exciton interaction strength from Eq.~(\ref{InteractionContact}), and $X_0$, $C_0$ are the Hopfield coefficients evaluated at $P=0$. We plot the resulting collective excitation spectrum in Figure~\ref{CollectiveExcitations} for GaAs quantum wells in a GaAs microcavity. While the Bogoliubov treatment of multi-component condensates is a well-established formalism, its application to the coupled LP–UP polariton system introduces a qualitatively new degree of tunability: both the effective masses and the interaction constants depend explicitly on the Hopfield coefficients. This dependence enables continuous tuning of the condensate’s collective properties through the cavity detuning and Rabi coupling, linking the underlying exciton–photon composition to experimentally observable superfluid characteristics.

\noindent
Eq.~(\ref{eq:collective_spectrum_hopfield}) describes the collective excitation spectrum in the system with coupled lower- and upper-polariton condensates within the mean-field framework. The interaction coefficients satisfy $G_{\mathrm{LP\!-\!UP}}^{2}=G_{\mathrm{LP}}G_{\mathrm{UP}}$, and the $2\times2$ interaction matrix becomes degenerate, so that the interaction matrix effectively reduces to a single dominant term. Physically, this condition implies that the two branches predominantly share a common total-density response, while their relative-density channel becomes weakly constrained. Within this simplified equilibrium treatment, the coupled condensates thus support one clearly defined linear (sound-like) excitation associated with the broken global $\mathrm{U}(1)$ symmetry and a second, low-energy mode that is only weakly dispersive. As shown in Figure~\ref{CollectiveExcitations}, the $j=1$ branch corresponds to the collective sound-like mode with velocity $c_1$ from Eq.~(\ref{SoundVelocityParametrized}), while the $j=2$ branch may be interpreted as a relative-phase–like fluctuation between the LP and UP components. We emphasize that identifying this second mode with a distinct symmetry-breaking channel would require a full time-dependent analysis of the coupled order parameters. Such a treatment goes beyond the present static mean-field framework, which only captures equilibrium density responses. In this limit, the second branch should therefore be regarded as a soft collective excitation originating from interbranch coupling, rather than as a rigorously established Goldstone or Leggett mode \cite{Wouters2007, Leggett2001}. In the long-wavelength limit $P \to 0$, the excitation spectrum becomes linear, $\Omega_{j,P} \approx c_j |P|$, with sound velocities

\begin{equation}
c_j = \sqrt{\frac{G_{\text{LP}}}{2 m_{\text{LP}}} + \frac{G_{\text{UP}}}{2 m_{\text{UP}}} 
+ (-1)^{j-1} \sqrt{\left(\frac{G_{\text{LP}}}{2 m_{\text{LP}}} - \frac{G_{\text{UP}}}{2 m_{\text{UP}}}\right)^2 + \frac{G_{\text{LP-UP}}^2}{m_{\text{LP}} m_{\text{UP}}}} } \ ,
\label{eq:sound_velocity_hopfield}
\end{equation}
where $m_{\text{LP}}$ and $m_{\text{UP}}$ are the effective masses of the lower and upper polaritons, respectively. Eq.~(\ref{eq:sound_velocity_hopfield}) shows how the Hopfield coefficients enter the collective spectrum through the renormalized couplings $G_{\text{LP}}, G_{\text{UP}},$ and $G_{\text{LP-UP}}$. This formulation preserves the full detuning and Rabi splitting dependence of the excitonic and photonic fractions. 

\medskip
\par\noindent
The parameterization of the condensate populations requires some care. 
The Hopfield coefficients \(|X|^2\) and \(|C|^2\) characterize the exciton–photon composition of an individual polariton mode, 
but they do not determine how the total polariton density is distributed between the lower (LP) and upper (UP) polariton branches. 
In realistic microcavity systems, this population balance is set by kinetic and reservoir processes—such as pumping, relaxation, and decay—or can be directly inferred from experiment. 
To capture this generally, we introduce a phenomologically motivated dimensionless population-split parameter \(\alpha \in [0,1]\), 
which specifies the fraction of the total polariton density \(n_0\) residing in the LP and UP branches as follows,
\begin{equation}
n_{\mathrm{LP},0} = \alpha n_0, \qquad n_{\mathrm{UP},0} = (1-\alpha)n_0.
\end{equation}

\noindent
Once \(\alpha\) (or equivalently \(n_{\mathrm{LP},0}\) and \(n_{\mathrm{UP},0}\)) is specified, the Bogoliubov analysis proceeds self-consistently to yield the excitation spectrum, sound velocity, and critical temperature.

\medskip
\noindent
Our Bogoliubov analysis remains valid for any externally supplied $(n_{\text{LP},0}, n_{\text{UP},0})$ from experiment or kinetic modeling. The subsequent sections explore how $c_s$ and $T_c$ depend on $\alpha$, 
providing testable signatures of two-component condensation.

\section{Tunable sound velocity}
\label{sec4}

\medskip
\par \noindent
Evaluating the expression for the sound-like branch ($j= 1$ branch), and expanding, gives the effective sound velocity
\begin{equation}
\boxed{
c_s = \sqrt{U_0\!\left(\frac{|X_0|^4 n_{\mathrm{LP},0}}{m_{\mathrm{LP}}}
+ \frac{|C_0|^4 n_{\mathrm{UP},0}}{m_{\mathrm{UP}}}\right)} }.
\label{SoundVelocityParametrized}
\end{equation}

\medskip
\par \noindent
This form makes explicit how the sound velocity depends on the Hopfield coefficients, the effective masses of the polariton branches, and the population imbalance between them. We begin by visualising how the sound velocity as a function of detuning for various values of the population split parameter.  In this work, the single-branch limits ($\alpha$~=~1 for LP-only and  $\alpha$= 0 for UP-only) are modeled as realistic polariton condensates confined to one branch, retaining the Hopfield-dependent effective masses and interaction constants.

\begin{figure}[H]
    \centering
    \subfigure[]{
        \includegraphics[width=0.5\textwidth]{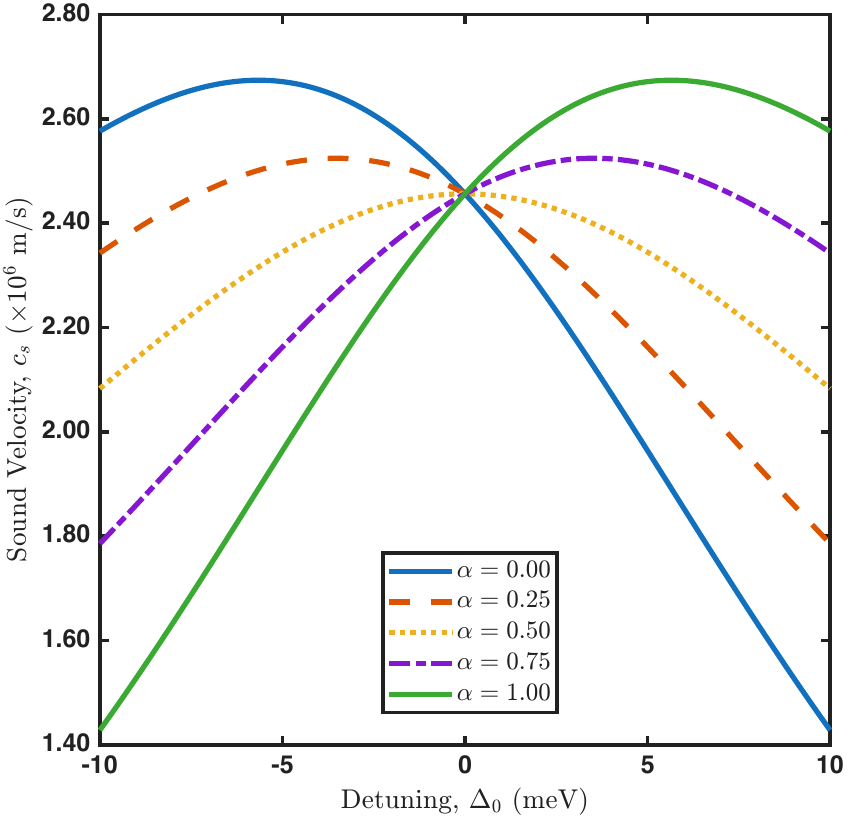}
    \label{fig:SoundVelocityVersusDetuning}
    }
    \hfill
    \subfigure[]{
        \includegraphics[width=0.45\textwidth]{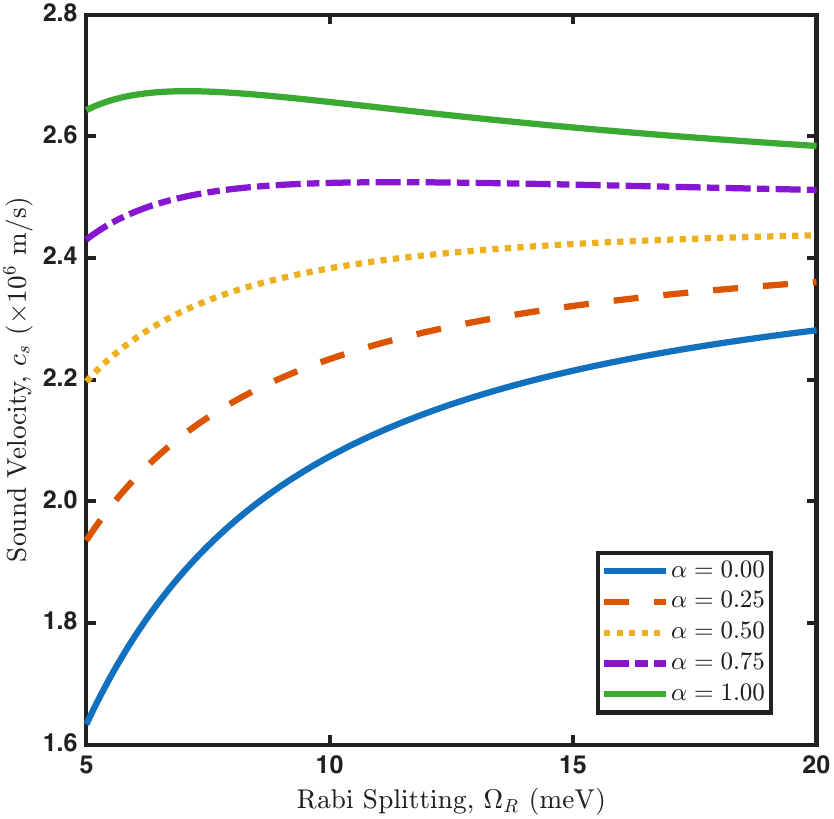}
        \label{fig:SoundVelocityVersusRabi}
    }
    \caption{Comparison of sound velocity $c_s$ in a two-component polariton condensate:
    (a) dependence on detuning $\Delta_0$ at fixed $\Omega_R = 8$ meV, and 
    (b) dependence on Rabi splitting $\Omega_R$ at fixed $\Delta_0 = 5$ meV. 
    In both cases the polariton density is $n = 1 \times 10^{11}~\text{cm}^{-2}$.}
    \label{fig:SoundVelocityDetuningAndRabi}
\end{figure}

\begin{figure}[H]
    \centering
    \subfigure[]{
        \includegraphics[width=0.45\textwidth]{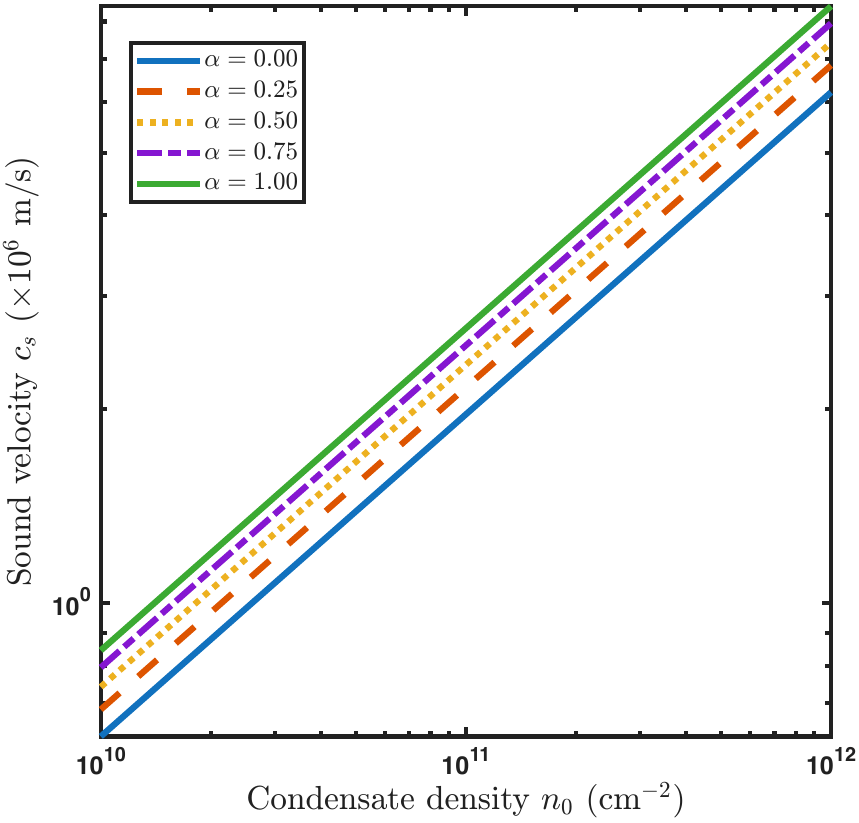}
        \label{fig:SoundVelocityVersusConc}
    }
    \hfill
    \subfigure[]{
        \includegraphics[width=0.45\textwidth]{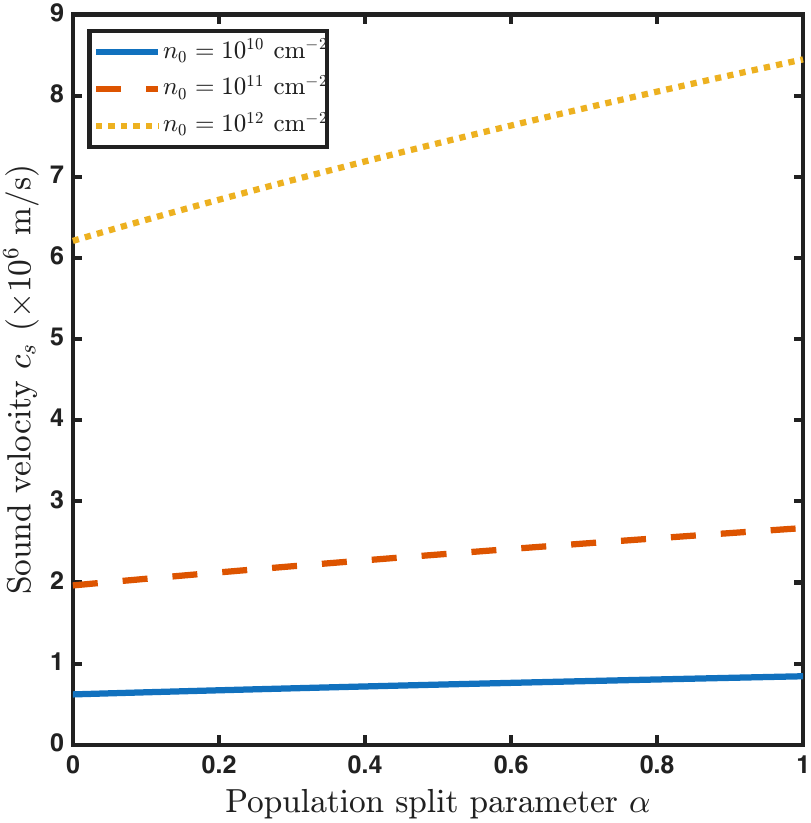}
        \label{fig:SoundVelocityVersusAlpha}
    }
    \caption{Sound velocity $c_s$ in a two-component polariton condensate as (a) a function of the total polariton density $n_0$ for different values of the condensate population--split parameter $\alpha$, (b) and as a function of the condensate population--split parameter $\alpha$ for various fixed total polariton densities. The Rabi splitting is fixed at $\Omega_R = 8~\text{meV}$ and the detuning at $\Delta_0 = 5~\text{meV}$.}
    \label{fig:SoundVelocityConcAndAlpha}
\end{figure}

At zero detuning ($\Delta_0 = 0$), the lower and upper polariton fractions are equal ($|X_0|^2 = |C_0|^2 = 1/2$), so their contributions to Eq.~(\ref{SoundVelocityParametrized}) balance, and the sound velocity converges to a single value for all population splits $\alpha$. This is because at zero detuning, the lower polariton and upper polariton effective masses, $m_{\text{LP}}$ and $m_{\text{UP}}$, stated analytically in Eq.
(\ref{LowerPolMass})-(\ref{UpperPolMass}), are equal. Away from resonance ($\Delta_0 \neq 0$), the Hopfield coefficients $|X_0|^2$ and $|C_0|^2$ become asymmetric, and the relative condensate populations shift the overall mean-field energy. As shown in Figure~\ref{fig:SoundVelocityDetuningAndRabi}(a), this causes the sound velocity to peak at different detunings depending on $\alpha$. For instance, when the condensate is predominantly lower polaritonic ($\alpha = 0.75$), the stronger excitonic character ($|X_0|^4$ term) enhances the effective interaction, leading to a higher $c_s$ that peaks near $\Delta_0 \!\approx\! 5~\mathrm{meV}$. Increasing the bare Rabi splitting leads to higher sound velocities, reflected in larger $c_s$ at fixed $\Omega_R$ for greater $\alpha$. The figure illustrates this trend for discrete coupling strengths and is not intended to imply a continuous variation of $\Omega_R$. In Figure~\ref{fig:SoundVelocityDetuningAndRabi}(b), a slightly positive detuning was chosen, so condensates with a greater lower-polariton fraction exhibit higher sound velocities, reflecting their stronger excitonic contribution to the interaction energy. We would also like to probe the sound velocity as a function of concentration for various total fixed polariton densities $n_0$ and $\alpha$ values in  Figure~(\ref{fig:SoundVelocityConcAndAlpha}).

The dependence of sound velocity on polariton density is shown in Figure~(\ref{fig:SoundVelocityConcAndAlpha})(a), where $c_s$ is plotted against $n_0$ on logarithmic scales for several values of the population-split parameter $\alpha$. The linear behavior in this log-log representation confirms the $c_s \propto \sqrt{n_0}$ scaling predicted by Eq.~(\ref{SoundVelocityParametrized}). At fixed density, condensates with higher LP fractions ($\alpha \to 1$) exhibit larger sound velocities due to the $|X_0|^4$ weighting of the LP self-interaction term, which is maximal at positive detunings where the LP branch is predominantly excitonic. Figure~\ref{fig:SoundVelocityConcAndAlpha}(b) shows the complementary view of the sound velocity $c_s$ versus $\alpha$ at fixed total densities $n_0$. At the chosen positive detuning ($\Delta_0 = 5~\mathrm{meV}$), the sound velocity grows monotonically with $\alpha$, interpolating smoothly between the pure UP limit and the pure LP limit for $n_0 = 1 \times 10^{11}~\mathrm{cm}^{-2}$. This monotonic $\alpha$-dependence is characteristic of positive detunings, where the lower polariton has a larger excitonic component. For negative detunings, the situation is reversed: the upper polariton becomes more excitonic, leading to a corresponding decrease in $c_s$ with increasing $\alpha$.

\section{Superfluidity}
\label{sec5}

\medskip
\par
\noindent
In this section, we determine the criterion for which a weakly interacting Bose gas of two components (upper and lower polaritons) can form a superfluid. We exploit the results from the preceding Section \ref{sec4}. Since at low momenta the energy spectrum of quasiparticles of a weakly interacting gas of polaritons is soundlike, this system satisfies the Landau criterion for superfluidity \cite{LifPit}. The critical velocity for superfluidity is given by $v_c = \text{min}(c_1,c_2)$ since the quasiparticles are created at velocities exceeding that of sound for the lowest mode of the quasiparticle dispersion.

\medskip
\par
\noindent
The density $\rho_s(T)$ of the superfluid is defined as $\rho_s(T) = \rho - \rho_n(T)$, where $\rho = m_{\text{UP}} n_{\text{UP},0} + m _{\text{LP}} n_{\text{LP},0}$ is the total 2D concentration of the system. Also,  $\rho_n(T)$ is the density of the normal component. We define the normal component $\rho_n(T)$ in the usual way \cite{berman2022nanomaterials}. We assume that the polariton system moves with velocity u, which means that the superfluid component has velocity u. At finite temperatures $T$, dissipating quasiparticles will emerge in this system. Since their density is small at low temperatures, one may assume that the gas of quasiparticles is an ideal Bose gas. We obtain the density of the superfluid component in the system following the procedure described in Ref. \cite{phosphoreneOGR}. To calculate the superfluid component density, we begin by defining the mass current J for a Bose gas for quasiparticles in the frame of reference where the superfluid component is at rest by

\begin{equation}
J =  \int \frac{d^2P}{(2 \pi \hbar)^2} P\left[f[\Omega_{1}(P) -Pu]+f[\Omega_{2}(P)-P\cdot]\right] \ ,
\label{massCurrent}
\end{equation}
where $f[\Omega_{1}(P)] = \{ \text{exp}[\Omega_{1}(P)/(k_B T)] -1\}^{-1} $ and $f[\Omega_{2}(P)] = \{ \text{exp}[\Omega_2(P)/(k_B T)] -1\}^{-1}$ are the Bose-Einstein distribution functions for the quasiparticles with the angle dependent dispersions $\Omega_{1}(P)$ and $\Omega_{2}(P)$ respectively. Expanding the expression under the integral in terms of $Pu/k_B T$ and restricting ourselves to the first-order term, we obtain

\begin{equation}
J = -\frac{u}{2} \int \frac{d^2P}{(2 \pi \hbar)^2} P^2 \left(\frac{\partial f [\Omega_{1}(P)]}{\partial \varepsilon_{1}(P)} +\frac{\partial f [\Omega_{2}(P)]}{\partial \Omega_{2}(P)}\right) \ .
\label{MassCurrentFirstOrder}
\end{equation}
\medskip
\par
\noindent
Using Eqs.~(\ref{massCurrent}) and (\ref{MassCurrentFirstOrder}), along with the definition $J = \rho_n u$, we define

\begin{equation}
\rho_n(T) = -\frac{1}{2} \int \frac{d^2P}{(2 \pi \hbar)^2} P^2 \left(\frac{\partial f [\Omega_{1}(P)]}{\partial \Omega_{1}(P)} +\frac{\partial f [\Omega_{2}(P)]}{\partial \Omega_{2}(P)}\right) \ .
\label{DensityFirstOrder}
\end{equation}

\medskip
\par
\noindent
Exploiting the methods described in Ref.~\cite{HighTempSuperOleg} for the low temperature case, we obtain the mean field critical temperature

\begin{equation}
T_c = \left[\frac{2 \pi \hbar^2 \rho c_s^4}{3 \zeta(3)  k_B^3}\right]^{1/3} \ ,
\label{critTempSoundVelocity}
\end{equation}
\noindent
where \( T_c\) is the critical temperature for superfluidity, \( \hbar \) is the reduced Planck's constant, \( \rho \) is the total 2D density of the system, accounting for contributions from both polariton branches, \( c_s\) is the sound velocity defined for the two-component condensate in Eq.~(\ref{SoundVelocityParametrized}) and for the one-component condensate are the limits $\alpha = 1$ and $\alpha = 0$. Additionally, \( \zeta(3) \) is the Riemann zeta function evaluated for a value of $3$,  and \( k_B \) is the Boltzmann constant. For the two component condensate of upper and lower polaritons, this is explicitly written in terms of the Hopfield coefficients and effective masses of the upper and lower polartions as follows

\begin{equation}
T_c  = \left[\frac{2 \pi \hbar^2 \rho}{3 \zeta(3)  k_B^3}\right]^{1/3} \left(U_0{\left( \frac{|X_0|^4\, n_{\mathrm{LP},0}}{m_{\mathrm{LP}}}
\;+\; \frac{|C_0|^4\, n_{\mathrm{UP},0}}{m_{\mathrm{UP}}} \right) }\right)^{2/3}.
\label{CritTemp}
\end{equation}

\noindent
We plot the dependance of the critical temperature from Eq.~(\ref{CritTemp}) as a function of the detuning, and Rabi splitting.

\begin{figure}[H]
    \centering
    \subfigure[]{
        \includegraphics[width=0.45\textwidth]{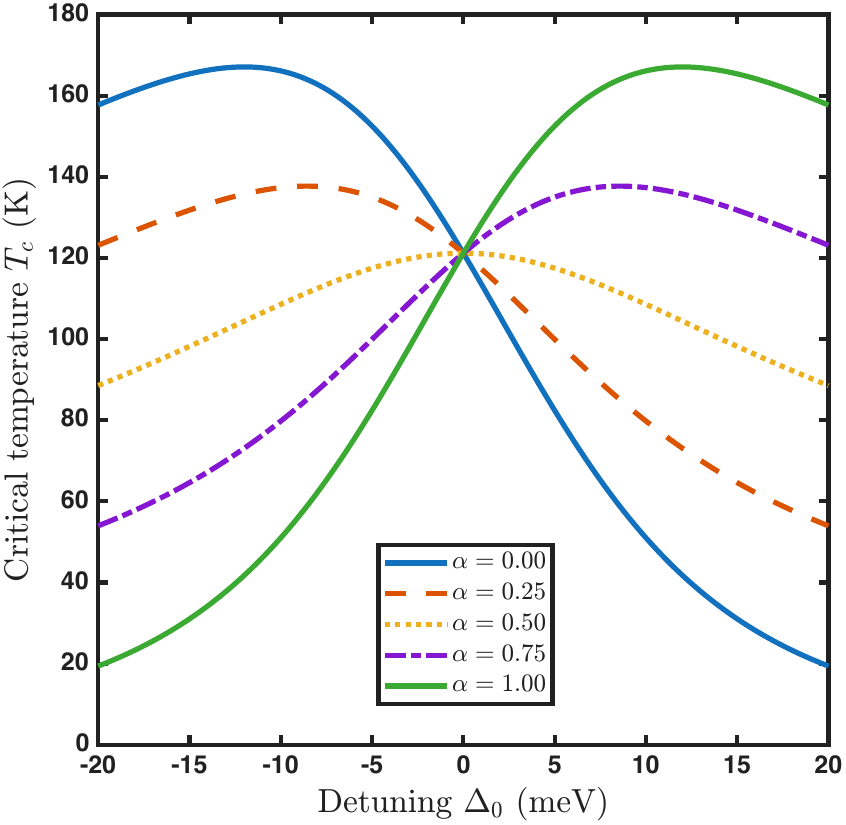}
        \label{fig:TempVsDetuning}
    }
    \hfill
    \subfigure[]{
        \includegraphics[width=0.45\textwidth]{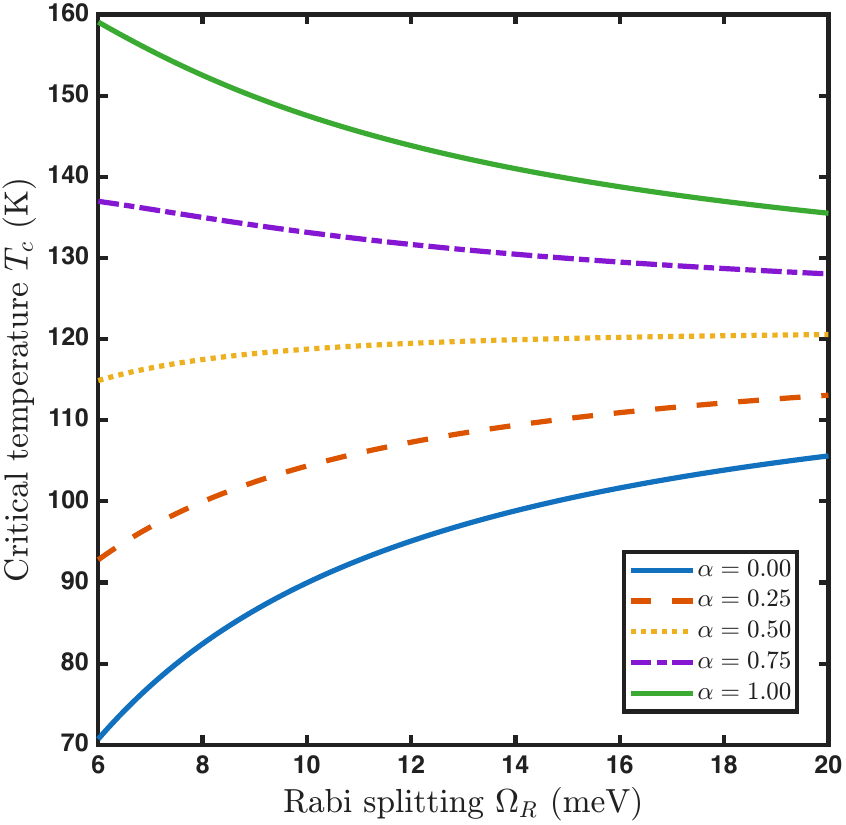}
        \label{fig:TempVsRabi}
    }
    \subfigure[]{
        \includegraphics[width=0.45\textwidth]{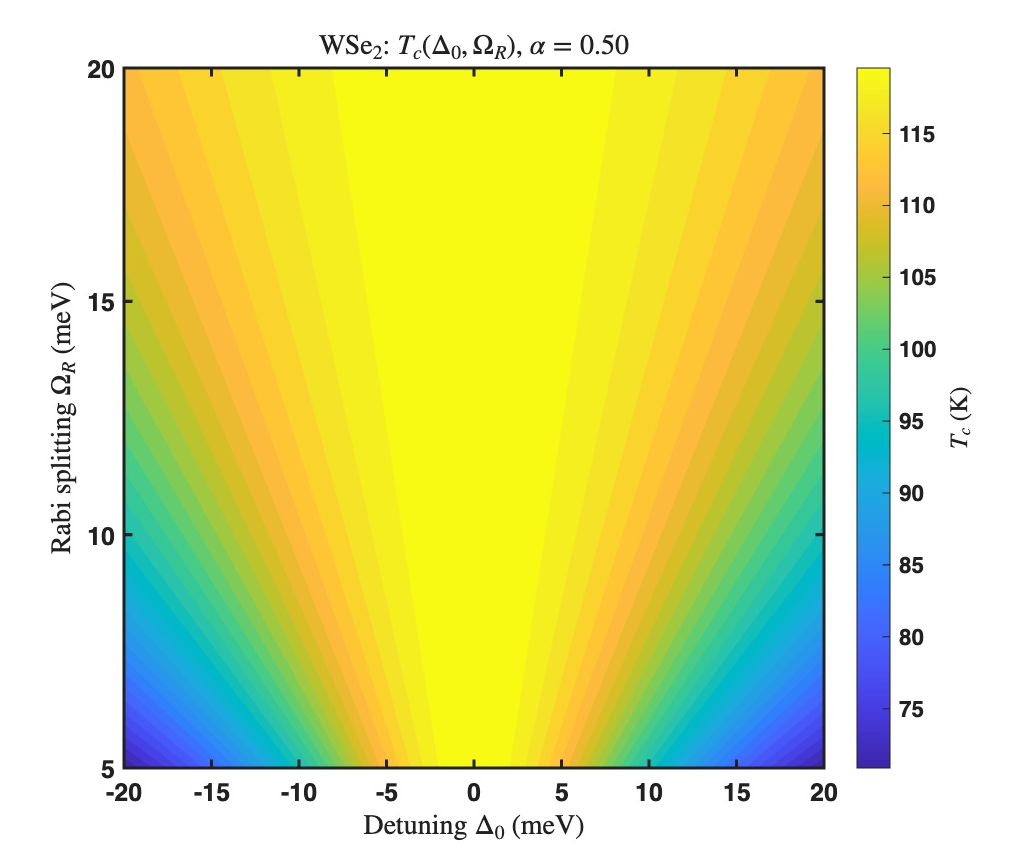}
        \label{fig:contourDetRabi}
    }
     \subfigure[]{
        \includegraphics[width=0.45\textwidth]{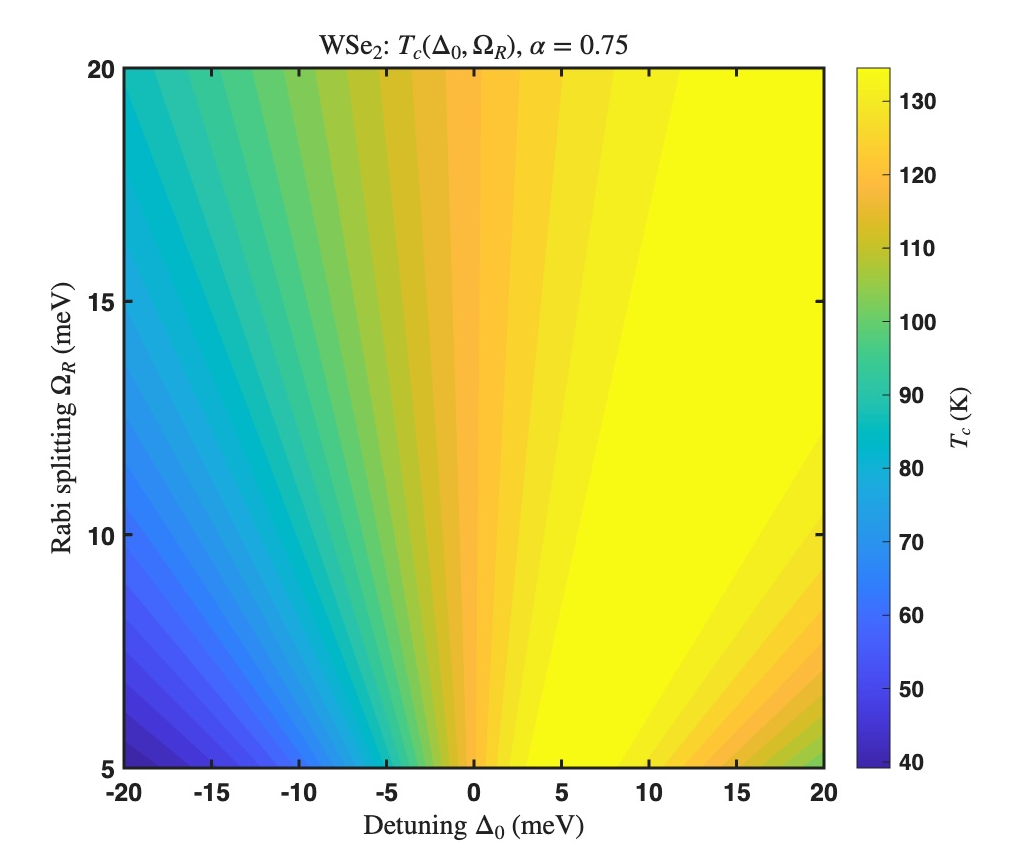}
        \label{fig:contourDetRabi0P75}
    }
    \caption{Critical temperature $T_c$ in a two-component polariton condensate in monolayer WSe$_2$. 
    (a) Dependence on detuning $\Delta_0$ at fixed Rabi splitting $\Omega_R = 8$ meV. 
    (b) Dependence on Rabi splitting $\Omega_R$ at fixed detuning $\Delta_0 = 5$ meV. 
    In both panels, the total polariton density is held constant at  $n_0 = 1 \times 10^{11}~\text{cm}^{-2}$, and results are shown for multiple values 
    of the population--split parameter $\alpha$. 
    (c) Contour map of $T_c$ as a function of $\Delta_0$ and $\Omega_R$ at fixed density 
    $n_0 = 1 \times 10^{11}~\text{cm}^{-2}$ and population split $\alpha=0.5$ and (d) $\alpha=0.75$. 
    This panel highlights how cavity detuning and light--matter coupling together shape 
    the critical temperature landscape.}
    \label{fig:CriticalTempDetuningAndRabi}
\end{figure}

At zero detuning ($\Delta_0 = 0$), the lower and upper polariton fractions are equal ($|X_0|^2 = |C_0|^2 = 1/2$), so their contributions to Eq.~(\ref{CritTemp}) balance, and the critical temperature converges to a single value for all population splits $\alpha$. Away from resonance ($\Delta_0 \neq 0$), the Hopfield coefficients $|X_0|^2$ and $|C_0|^2$ become asymmetric, and the relative condensate populations shift both the sound velocity entering Eq.~(\ref{CritTemp}) and the 2D total concentration of the system~$\rho$. As shown in Figure~\ref{fig:CriticalTempDetuningAndRabi}(a), this causes $T_c$ to peak at different detunings depending on $\alpha$. For instance, when the system is predominantly lower polaritonic ($\alpha = 0.75$), the enhanced excitonic character increases the interaction-driven sound velocity, which enters as $c_s^{4/3}$ in the mean-field expression for $T_c$, leading to a higher critical temperature that peaks near $\Delta_0 \!\approx\! 10~\mathrm{meV}$. Increasing the Rabi splitting reduces $T_c$ initially for all $\alpha$, before levelling out, with the effect most pronounced for larger values of $\alpha$ where the excitonic contribution to the interaction energy is strongest. In Figure~\ref{fig:CriticalTempDetuningAndRabi}, a slightly positive detuning was chosen, so condensates with a greater lower-polariton fraction exhibit higher critical temperatures, reflecting the combined effect of stronger interaction and larger effective mass in determining the superfluid stability.

\medskip
\par
\noindent
In addition to the critical temperature, it is convenient to define a critical polariton concentration $n_c$ at a given system temperature $T$.  This quantity characterizes the minimal density required for superfluidity to emerge at temperature $T$ within our mean-field framework. Starting from the expression for the critical temperature,
\begin{equation}
T_c  = \left[\frac{2 \pi \hbar^2 \rho}{3 \zeta(3)  k_B^3}\right]^{1/3} 
\left[ U_0 \left( \frac{|X_0|^4\, n_{\mathrm{LP},0}}{m_{\mathrm{LP}}}
+ \frac{|C_0|^4\, n_{\mathrm{UP},0}}{m_{\mathrm{UP}}} \right) \right]^{2/3},
\label{CritTemp}
\end{equation}
we substitute the definitions 
$n_{\mathrm{LP},0} = \alpha n_0$ and $n_{\mathrm{UP},0} = (1-\alpha)n_0$, 
and introduce the notation
\begin{equation}
S(\Delta_0,\Omega_R,\alpha) = 
\frac{|X_0|^4}{m_{\mathrm{LP}}}\,\alpha 
+ \frac{|C_0|^4}{m_{\mathrm{UP}}}\,(1-\alpha).
\label{Sfactor}
\end{equation}
Eq.~(\ref{CritTemp}) can then be written as
\begin{equation}
T_c^3 = \frac{2\pi \hbar^2 \rho}{3 \zeta(3) k_B^3} \,
\big( U_0 S n_0 \big)^2.
\end{equation}
\noindent
We also note that the density is defined as, 
 
\begin{equation}
\rho = m_{\mathrm{UP}} n_{\mathrm{UP},0} + m_{\mathrm{LP}} n_{\mathrm{LP},0}
= n_0 \Big[ m_{\mathrm{LP}} \alpha + m_{\mathrm{UP}} (1-\alpha) \Big] 
\equiv n_0 M_\alpha,
\label{rhoMassDensity}
\end{equation}
where $M_\alpha$ is a weighted polariton effective mass determined by the population 
split parameter $\alpha$. Substituting Eq.~(\ref{rhoMassDensity}) for $\rho$ gives
\begin{equation}
T_c^3 = \frac{2\pi \hbar^2}{3 \zeta(3) k_B^3}\,
M_\alpha \,(U_0 S)^2\, n_0^3.
\end{equation}
Defining
\begin{equation}
A \equiv \frac{2\pi \hbar^2}{3 \zeta(3) k_B^3},
\end{equation}
we may write
\begin{equation}
{
T_c(\Delta_0,\Omega_R,n_0,\alpha) 
= \big( A M_\alpha \big)^{1/3}\,(U_0 S)^{2/3}\,n_0
}
\label{TcFinal}
\end{equation}
which shows that $T_c$ is {linear} in the total polariton concentration $n_0$ 
once the density-dependence of $\rho$ is taken into account. Finally, inverting Eq.~(\ref{TcFinal}) allows us to identify the critical 
concentration $n_c$ at a given temperature $T$ as
\begin{equation}
\boxed{ \;
n_c(T;\Delta_0,\Omega_R,\alpha) =
\frac{T}{\big( A M_\alpha \big)^{1/3}\,(U_0 S)^{2/3}}
\;}.
\label{CritConcFinal}
\end{equation}

\medskip
\par
\noindent
Both $S$ and $M_\alpha$ encode the $\alpha$-dependence of the problem. In particular
$S$ does so through the Hopfield coefficients and effective polariton masses, and $M_\alpha$ through the mass-weighted population imbalance. Thus, measurements of either $T_c$ or $n_c$ provide a direct probe of the relative occupations of the condensates in the lower and upper polariton branches. The critical concentration $n_c$ provides an experimentally accessible quantity that complements the critical temperature $T_c$. At fixed detuning $\Delta_0$, Rabi splitting $\Omega_R$, and interaction strength $U_0$, our mean-field framework predicts a one-to-one mapping between $n_c$ and the condensate population-split parameter $\alpha$.
\medskip
\par
\noindent
This formulation suggests a concrete experimental strategy by measuring the critical density at which superfluidity onsets at a known temperature, an experimentalist can read off the corresponding value of $\alpha$ from our theoretical curves plotted below in Figure~(\ref{fig:AlphaVsNcAndTc}). This procedure yields two pieces of information simultaneously. The value of $\alpha$, quantifying the relative occupations of the condensates in the lower and upper polariton branches. Additionally it gives confirmation of whether the system indeed realizes a genuine two-component condensate. If the extracted $\alpha$ lies away from the single-component limits ($\alpha=0$ or $\alpha=1$), this serves as direct evidence of coexistence of both branches. Thus, the dependence of $n_c$ on $\alpha$ transforms the critical concentration into an experimental probe of population imbalance, offering a pathway to distinguish two-component polariton condensation from its single-component counterparts. An entirely analogous procedure can be applied to the critical temperature $T_c$. Measuring $T_c$ at fixed density allows one to extract the underlying population imbalance in the same manner.

\begin{figure}[H]
    \centering
    \subfigure[]{
        \includegraphics[width=0.45\textwidth]{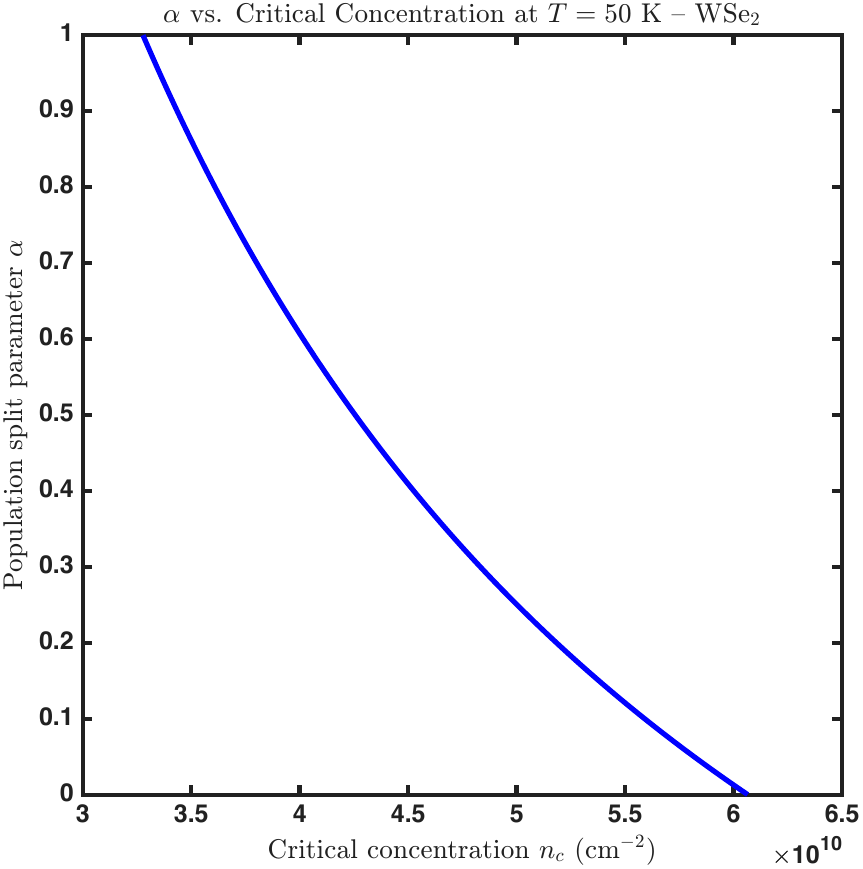}
        \label{fig:AlphaVsNc}
    }
    \hfill
    \subfigure[]{
        \includegraphics[width=0.45\textwidth]{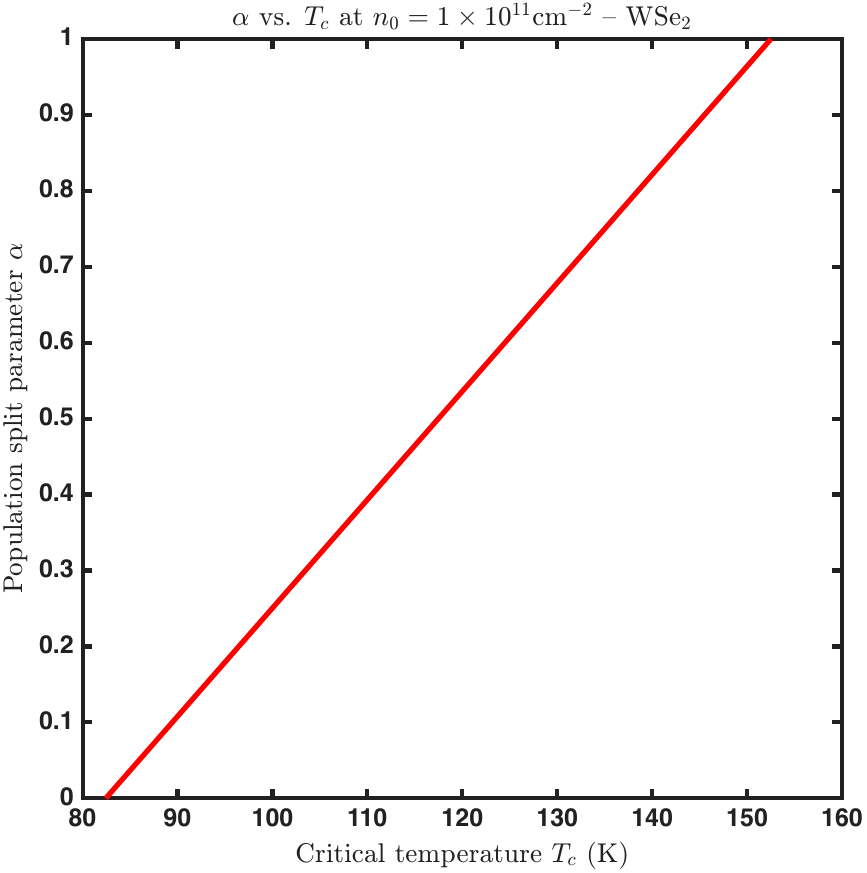}
        \label{fig:AlphaVsTc}
    }
    \caption{Condensate population--split parameter $\alpha$ in a two-component polariton condensate in WSe$_2$ monolayers embedded in a GaAs microcavity. 
    (a) Dependence of $\alpha$ on the critical concentration $n_c$ at fixed system temperature $T=50$~K, Rabi splitting $\Omega_R = 8$~meV, and detuning $\Delta_0 = 5$~meV. 
    (b) Dependence of $\alpha$ on the critical temperature $T_c$ at fixed total polariton density $n_0 = 1 \times 10^{11}~\text{cm}^{-2}$, Rabi splitting $\Omega_R = 8$~meV, and detuning $\Delta_0 = 10$~meV.}
    \label{fig:AlphaVsNcAndTc}
\end{figure}

\section{Discussion and conclusions}
\label{conclusion}

\medskip
\par
\noindent
We have developed a microscopic mean-field theory of two-component superfluidity in semiconductor microcavities, treating coexisting condensates of lower and upper polaritons as coupled components of a unified superfluid, with their distinct dispersions, effective masses, and interaction strengths emerging from the underlying exciton-photon Hamiltonian. By incorporating the full momentum and detuning dependence of the Hopfield coefficients, effective masses, and interaction strengths, we derived explicit expressions for the collective excitation spectrum, sound velocity $c_s$, and critical temperature $T_c$ as functions of the cavity detuning $\Delta_0$, Rabi splitting $\Omega_R$, and our newly introduced, phenomenological population-split parameter $\alpha$. Our central result is that two-component condensation produces experimentally measurable signatures that distinguish it from conventional single-branch superfluidity, that is enhanced sound velocity and critical temperature across most of the parameter space, distinctive detuning-dependent peak structures in $c_s(\Delta_0,\Omega_R)$, $T_c(\Delta_0,\Omega_R)$, and $n_c (\Delta_0, \Omega_R)$ and a systematic dependence on $\alpha$ that enables direct experimental extraction of the population split from measurements at known system parameters.

\medskip
\par
\noindent
At zero detuning, where $|X_0|^2 = |C_0|^2 = 1/2$, the system is symmetric and $c_s$ and $T_c$ converge to single values independent of $\alpha$. Away from resonance, the Hopfield-weighted interaction terms $g_{\ell\ell} = U_0|X_0|^4$, $g_{uu} = U_0|C_0|^4$, and $g_{\ell u} = U_0|X_0|^2|C_0|^2$ become asymmetric, and both $c_s$ and $T_c$ develop pronounced $\alpha$-dependence. For instance, at positive detunings in monolayer WSe$_2$ ($\Delta_0 = 5~\mathrm{meV}$, $\Omega_R = 8~\mathrm{meV}$), an LP-dominated condensate ($\alpha = 0.75$) reaches a peak critical temperature  $T_c \approx 140~\mathrm{K}$, compared to  $T_c \approx 120~\mathrm{K}$ for a balanced condensate ($\alpha = 0.5$). These enhancements arise from the stronger excitonic contribution to the interaction energy when the LP branch is predominantly excitonic in nature.

\medskip
\par
\noindent
Many examples demonstrate that multi-component condensation with interbranch coherence is not merely hypothetical but has been realized experimentally in systems with comparable energy splittings and interaction-driven coupling. The magnon case is particularly relevant: Bailey et al. (2024) \cite{Aeppli2022} report multi-band Bose-Einstein condensation stabilized by four-particle scattering processes, demonstrating that interaction-mediated coupling can sustain coexisting populations in energetically distinct branches. This experiment was performed in thin films of Yttrium Iron Garnet (YIG), where quantized spin waves (magnons) form quantized subbands due to confinement across the film thickness. Under microwave pumping, Bose–Einstein condensation occurring simultaneously in two distinct magnon bands has been observed. One corresponds to the lowest (Kittel-like) mode and the other to an excited quantized mode. The coexistence of these two condensates is stabilized by strong four-magnon scattering processes that exchange pairs of magnons between the bands, effectively coupling the condensates despite their energy splitting. This establishes a genuine multi-band condensate with interaction-mediated coherence between energetically distinct branches. It is directly analogous to our LP–UP system, where interbranch scattering terms in Eq. (15) play the same coupling role, suggesting that two-component condensation stabilized by interaction-driven coupling is experimentally realistic.

\medskip
\par
\noindent
Crucially, our framework is intentionally predictive rather than explanatory. We do not derive the microscopic conditions under which simultaneous LP and UP condensation occurs—that question requires a full kinetic theory incorporating reservoir dynamics, phonon scattering, pump mechanisms, and the conditions for phase-locking between branches—and remains an important target for future work. Instead, we establish theoretical benchmarks that answer the question: \textit{if} both branches are macroscopically occupied, \textit{what} collective phenomena emerge, and \textit{how} can they be unambiguously identified?
\medskip
\par
\noindent
Figure~\ref{fig:AlphaVsNcAndTc} provide explicit extraction protocols in that by measuring $c_s$ or $T_c$ at known $n_0$ and $\Delta_0$ and $\Omega_R$, experimentalists can infer $\alpha$ and thereby confirm the presence and degree of two-component condensation. These predictions are general and apply to any strongly coupled exciton-photon system, with numerical parameters adjusted for specific materials (GaAs/AlGaAs quantum wells and monolayer TMDCs embedded in GaAs microcavities).

\medskip
\par
\noindent
While most experiments to date observe single-branch condensation due to efficient phonon-assisted relaxation, recent demonstrations of upper-polariton condensation~\cite{Chen2023}, suggest that controlled dual-branch occupation may be experimentally accessible under appropriate conditions. These developments echo successful realizations of multi-component condensation in related systems: spinor atomic BECs exhibit simultaneous occupation of multiple hyperfine states with interaction-driven phase-locking~\cite{Stamper2013,Ho1998}, magnon condensates display coexisting branches with hybridized collective modes~\cite{Aeppli2022}. Our results provide the theoretical framework needed to recognize and characterize analogous phenomena when LP and UP branches are simultaneously macroscopically occupied.

\medskip
\par
\noindent
Looking forward, several natural extensions of this work merit investigation. A complete understanding of the coupled LP–UP condensate will ultimately require a time-dependent treatment of the relative phase between the two order parameters, which may give rise to Rabi-like oscillations or other dynamical coherence phenomena beyond the scope of the present equilibrium model. First, a driven–dissipative formulation based on the Lindblad master equation would capture the intrinsically nonequilibrium nature of polariton condensates, incorporating pumping, decay, and reservoir dynamics explicitly to provide time-dependent predictions for $\alpha(t)$ and to determine the conditions under which steady-state dual-branch occupation can emerge. Second, a full amplitude–phase decomposition of the coupled order parameters—together with a microscopic analysis of the phase-locking condition—remains an essential next step. Such a study would establish how interaction-induced coherence and fluctuations determine the stability of the relative phase, thereby connecting the phenomenological population-split parameter $\alpha$ to experimentally controllable pump and reservoir parameters. Finally, our framework naturally generalizes to other multi-component light–matter systems, including cavity magnon-polaritons, phonon-polaritons, and multi-mode photonic condensates, where interaction-driven coupling between branches may yield analogous collective phenomena. By formulating a quantitative equilibrium baseline for two-component polariton superfluidity, this work provides both conceptual motivation and measurable benchmarks for future microscopic and experimental investigations.

\section*{Acknowledgements}

A. Nafis Arafat  and Godfrey Gumbs gratefully acknowledge funding from the U.S. National Aeronautics and Space Administration (NASA) via the NASA-Hunter College Center for Advanced Energy Storage for Space  under cooperative agreement 80NSSC24M0177.  Oleg Berman acknowledges the support from PSC CUNY Award No. 66382-00 54. This research benefited from Physics Frontier Center for Living Systems funded by the National Science Foundation (PHY- 2317138). The authors are grateful to David W. Snoke  for
valuable discussions. 


\newpage 
\appendix
\section{Systematic interaction reduction}
\label{InteractionReduction}

We outline in this section how we systematically reduce Hamiltonian interaction terms to obtain the simplified form in Eq. (\ref{FinalHamiltonianKspace}). Let us begin with the 16 interaction terms

\begin{equation}
\begin{aligned}
\frac{1}{2A} \sum_{\substack{P, P', q}} U_q \Bigg[ 
& X_{P+q} X_{P'-q} X_P X_{P'} \hat{l}_{P+q}^\dagger \hat{l}_{P'-q}^\dagger \hat{l}_P \hat{l}_{P'} 
- X_{P+q} X_{P'-q} C_P X_{P'} \hat{l}_{P+q}^\dagger \hat{l}_{P'-q}^\dagger \hat{u}_P \hat{l}_{P'} \\
& - X_{P+q} X_{P'-q} X_P C_{P'} \hat{l}_{P+q}^\dagger \hat{l}_{P'-q}^\dagger \hat{l}_P \hat{u}_{P'} 
+ X_{P+q} X_{P'-q} C_P C_{P'} \hat{l}_{P+q}^\dagger \hat{l}_{P'-q}^\dagger \hat{u}_P \hat{u}_{P'} \\
& - X_{P+q} C_{P'-q} X_P X_{P'} \hat{l}_{P+q}^\dagger \hat{u}_{P'-q}^\dagger \hat{l}_P \hat{l}_{P'} 
+ X_{P+q} C_{P'-q} C_P X_{P'} \hat{l}_{P+q}^\dagger \hat{u}_{P'-q}^\dagger \hat{u}_P \hat{l}_{P'} \\
& - X_{P+q} C_{P'-q} X_P C_{P'} \hat{l}_{P+q}^\dagger \hat{u}_{P'-q}^\dagger \hat{l}_P \hat{u}_{P'} 
+ X_{P+q} C_{P'-q} C_P C_{P'} \hat{l}_{P+q}^\dagger \hat{u}_{P'-q}^\dagger \hat{u}_P \hat{u}_{P'} \\
& - C_{P+q} X_{P'-q} X_P X_{P'} \hat{u}_{P+q}^\dagger \hat{l}_{P'-q}^\dagger \hat{l}_P \hat{l}_{P'} 
+ C_{P+q} X_{P'-q} C_P X_{P'} \hat{u}_{P+q}^\dagger \hat{l}_{P'-q}^\dagger \hat{u}_P \hat{l}_{P'} \\
& - C_{P+q} X_{P'-q} X_P C_{P'} \hat{u}_{P+q}^\dagger \hat{l}_{P'-q}^\dagger \hat{l}_P \hat{u}_{P'} 
+ C_{P+q} X_{P'-q} C_P C_{P'} \hat{u}_{P+q}^\dagger \hat{l}_{P'-q}^\dagger \hat{u}_P \hat{u}_{P'} \\
& + C_{P+q} C_{P'-q} X_P X_{P'} \hat{u}_{P+q}^\dagger \hat{u}_{P'-q}^\dagger \hat{l}_P \hat{l}_{P'} 
- C_{P+q} C_{P'-q} C_P X_{P'} \hat{u}_{P+q}^\dagger \hat{u}_{P'-q}^\dagger \hat{u}_P \hat{l}_{P'} \\
& - C_{P+q} C_{P'-q} X_P C_{P'} \hat{u}_{P+q}^\dagger \hat{u}_{P'-q}^\dagger \hat{l}_P \hat{u}_{P'} 
+ C_{P+q} C_{P'-q} C_P C_{P'} \hat{u}_{P+q}^\dagger \hat{u}_{P'-q}^\dagger \hat{u}_P \hat{u}_{P'}
\Bigg].
\end{aligned}
\end{equation}

There are only two straightforward self-interaction terms for the upper and lower polaritons. These are 

\begin{equation}
X_{P+q} X_{P'-q} X_P X_{P'} \hat{l}_{P+q}^\dagger \hat{l}_{P'-q}^\dagger \hat{l}_P \hat{l}_{P'}
\label{self_lower}
\end{equation}

\begin{equation}
C_{P+q} C_{P'-q} C_P C_{P'} \hat{u}_{P+q}^\dagger \hat{u}_{P'-q}^\dagger \hat{u}_P \hat{u}_{P'}
\label{self_upper}
\end{equation}

\medskip
\par
\noindent
To simplify and rearrange the 16 interaction terms, we will utilize commutation relations for the bosonic creation and annihilation operators \cite{AGD2012}, as well as symmetry properties of the terms. 
The commutation relations reflect the nature of bosons, which can occupy the same quantum state without any restriction. The basic commutation relations for bosonic operators are,

\begin{equation}
[\hat{l}_P, \hat{l}_{P'}^\dagger] = \delta_{P,P'}
\end{equation}
\begin{equation}
[\hat{u}_P, \hat{u}_{P'}^\dagger] = \delta_{P,P'}
\end{equation}
\begin{equation}
[\hat{l}_P, \hat{l}_{P'}] = 0, \quad [\hat{u}_P, \hat{u}_{P'}] = 0
\end{equation}
where \( \hat{l}_P \) and \( \hat{u}_P \) are the annihilation operators for different polariton states. Additionally, since these operators correspond to different modes, their cross-commutators will also be zero:
\begin{equation}
[\hat{l}_P, \hat{u}_{P'}] = [\hat{l}_P, \hat{u}_{P'}^\dagger] = 0
\end{equation}

\medskip
\par
\noindent
Let us set aside the self interaction terms Eq. (\ref{self_lower})-(\ref{self_upper}) for now. We categorize the 14 remaining cross-interaction terms into two main groups based on the operator structures and then apply normal ordering where possible. Terms Involving \( \hat{l}^{\dagger} \hat{l} \hat{u}^{\dagger} \hat{u} \) or \( \hat{u}^{\dagger} \hat{u} \hat{l}^{\dagger} \hat{l} \) are interesting to us for considering physical scattering processes.

\medskip
\par
\noindent
The terms that fit this structure are
\begin{align*}
& X_{P+q} C_{P'-q} X_P C_{P'} \hat{l}_{P+q}^\dagger \hat{l}_P \hat{u}_{P'-q}^\dagger \hat{u}_{P'} \\
& + C_{P+q} X_{P'-q} C_P X_{P'} \hat{u}_{P+q}^\dagger \hat{u}_P \hat{l}_{P'-q}^\dagger \hat{l}_{P'}
\end{align*}

\medskip
\par
\noindent
These two terms involve interactions where two polaritons of different types (upper and lower) scatter with respect to each other. The terms that are remaining involve various other combinations of processes. These remaining 12 cross-interaction terms we categorize into two main groups based on the nature of the exchange processes: direct exchange terms and mixed exchange terms. The direct exchange terms represent direct interactions where the upper polariton exchanges with the lower polariton.

\begin{align}
& X_{P+q} C_{P'-q} C_P X_{P'} \, \hat{l}_{P+q}^\dagger \,\hat{u}_{P'-q}^\dagger \,\hat{u}_P \,\hat{l}_{P'} 
\nonumber \\[5pt]
&\quad + C_{P+q} X_{P'-q} X_P C_{P'} \,\hat{u}_{P+q}^\dagger \,\hat{l}_{P'-q}^\dagger \,\hat{l}_P \,\hat{u}_{P'} 
\label{directExchange}
\end{align}

\medskip
\par
\noindent
These two terms describe simple polariton exchange interactions. The following mixed terms involve more complex combinations of operators that we would like to sort through

\begin{align}
& X_{P+q} C_{P'-q} X_P X_{P'} \hat{l}_{P+q}^\dagger \hat{u}_{P'-q}^\dagger \hat{l}_P \hat{l}_{P'} 
\nonumber \\
& + X_{P+q} C_{P'-q} C_P C_{P'} \hat{l}_{P+q}^\dagger \hat{u}_{P'-q}^\dagger \hat{u}_P \hat{u}_{P'} 
\nonumber \\
& + C_{P+q} X_{P'-q} C_P C_{P'} \hat{u}_{P+q}^\dagger \hat{l}_{P'-q}^\dagger \hat{u}_P \hat{u}_{P'} 
\nonumber \\
& + C_{P+q} X_{P'-q} X_P X_{P'} \hat{u}_{P+q}^\dagger \hat{l}_{P'-q}^\dagger \hat{l}_P \hat{l}_{P'} 
\nonumber \\
& + X_{P+q} X_{P'-q} C_P X_{P'} \hat{l}_{P+q}^\dagger \hat{l}_{P'-q}^\dagger \hat{u}_P \hat{l}_{P'} 
\nonumber \\
& + X_{P+q} X_{P'-q} X_P C_{P'} \hat{l}_{P+q}^\dagger \hat{l}_{P'-q}^\dagger \hat{l}_P \hat{u}_{P'} 
\nonumber \\
& + C_{P+q} C_{P'-q} X_P C_{P'} \hat{u}_{P+q}^\dagger \hat{u}_{P'-q}^\dagger \hat{l}_P \hat{u}_{P'} 
\nonumber \\
& + C_{P+q} C_{P'-q} X_P X_{P'} \hat{u}_{P+q}^\dagger \hat{u}_{P'-q}^\dagger \hat{l}_P \hat{l}_{P'} 
\nonumber \\
& + X_{P+q} C_{P'-q} C_P X_{P'} \hat{l}_{P+q}^\dagger \hat{u}_{P'-q}^\dagger \hat{u}_P \hat{l}_{P'} 
\nonumber \\
& + C_{P+q} X_{P'-q} X_P C_{P'} \hat{u}_{P+q}^\dagger \hat{l}_{P'-q}^\dagger \hat{l}_P \hat{u}_{P'} 
\nonumber \\
& + X_{P+q} C_{P'-q} C_P X_{P'} \hat{l}_{P+q}^\dagger \hat{u}_{P'-q}^\dagger \hat{u}_P \hat{l}_{P'} 
\nonumber \\
& + C_{P+q} X_{P'-q} X_P C_{P'} \hat{u}_{P+q}^\dagger \hat{l}_{P'-q}^\dagger \hat{l}_P \hat{u}_{P'} 
\label{extraneous}
\end{align}

\medskip
\par
\noindent
{The terms in Eq. (\ref{extraneous}) involve three creation/annihilation operators of one type and one of the other. The interactions still involve four polaritons, but the arrangement is asymmetric with respect to ratio of upper and lower polaritons, and extremely unlikely to occur given the Rabi splitting.}

\section{Expanded Polariton Spectrum and Effective Masses}

Polaritons, which result from the strong coupling between excitons and photons, have an energy spectrum that reflects the combined properties of both components. To understand the behavior of polaritons, it is essential to compute their energy spectrum \( \varepsilon_{\text{LP/UP}}(P) \), where \( \text{LP} \) and \( \text{UP} \) refer to the lower and upper polariton branches, respectively. This spectrum depends on both the photon dispersion \( \varepsilon_{\text{ph}}(P) \) and the excitonic dispersion \( \varepsilon_{\text{exc}}(P) \).

\medskip
\par
\noindent
By analyzing the polariton spectrum in momentum space, we can extract the effective masses of the polaritons, which determine how they behave in real space under external potentials or forces. The effective mass \( m_{\text{eff}} \) for polaritons can be obtained by taking the second derivative of the energy spectrum \( \varepsilon_{\text{LP/UP}}(P) \) with respect to momentum \( P \):

\begin{equation}
m_{\text{eff}} = \left( \frac{\partial^2 \varepsilon_{\text{LP/UP}}(P)}{\partial P^2} \right)^{-1}
\end{equation}

\medskip
\par
\noindent
Here, \( \varepsilon_{\text{LP/UP}}(P) \) is the energy of the lower or upper polariton, and \( P \) is the momentum of the polariton. Once the effective mass is known, we can model the polaritons in real space. The polariton energy is given by

\begin{equation}
\varepsilon_{\text{LP/UP}}(P) = \frac{\varepsilon_{\text{ph}}(P) + \varepsilon_{\text{exc}}(P)}{2} \mp \frac{1}{2} \sqrt{\left( \varepsilon_{\text{ph}}(P) - \varepsilon_{\text{exc}}(P) \right)^2 + 4| \hbar \Omega_R^2|} \ ,
\label{UpLowS1}
\end{equation}

\noindent
Where, \( \varepsilon_{\text{LP}}(P) \) and \( \varepsilon_{\text{UP}}(P) \) are the lower and upper polariton energy dipsersions, respectively, \( \varepsilon_{\text{ph}}(P) \) is the photon energy dispersion, \( \varepsilon_{\text{exc}}(P) \) is the exciton energy dispersion, \( \Omega_R \) is the Rabi splitting, describing the coupling strength between the exciton and photon.
\medskip
\par
\noindent
For photons confined to a microcavity, the energy dispersion is given by
\begin{equation}
\varepsilon_{\text{ph}}(P) = \frac{c}{n} \sqrt{P^2 + \hbar^2 \pi^2 L_c^{-2}} \ .
\end{equation}

\medskip
\par
\noindent
Where, \( c \) is the speed of light in vacuum, \( n \) is the refractive index of the material inside the cavity, \( P \) is the in-plane momentum of the photon, \( \hbar \) is the reduced Planck's constant, \( L_c \) is the length of the cavity in the direction perpendicular to the plane, \( \hbar \pi L_c^{-1} \) represents the quantized photon momentum due to the confinement.

\medskip
\par
\noindent
The exciton energy dispersion is given by
\begin{equation}
\varepsilon_{\text{ex}}(P) = E_{\text{band}} - E_{\text{binding}} + \varepsilon_0(P) \ .
\label{PhotonDisp}
\end{equation}
Where, \( E_{\text{band}} \) is the band gap energy, \( E_{\text{binding}} \) is the exciton binding energy, \( \varepsilon_0(P) \) is the kinetic energy of the exciton, given by

    \begin{equation}
    \varepsilon_0(P) = \frac{P^2}{2 M_{\text{exc}}}
    \label{ExcitonDisp}
    \end{equation}

\medskip
\par
\noindent
\( M_{\text{exc}} \) is the effective mass of the exciton ($M_{\text{exc}} = m_e + m_h$), and \( P \) is the in-plane momentum of the exciton. The photon energy can be approximated to first-order as

\begin{equation}
\varepsilon_{\text{ph}}(P) \approx \frac{c}{n} \left( \hbar \pi L_c^{-1} + \frac{P^2}{2 \hbar \pi L_c^{-1}} \right)
\end{equation}

\medskip
\par
\noindent
Now we let \( \Delta_0 = \frac{c}{n} \hbar \pi L_c^{-1} - \left( E_{\text{band}} - E_{\text{binding}} \right) \).

\medskip
\par
\noindent
The expanded square root term is
\begin{equation}
\sqrt{\Delta_0^2 + 4 \Omega_R^2 + 2 \Delta_0 \frac{P^2}{2} \left( \frac{c}{n \hbar \pi L_c^{-1}} - \frac{1}{M_{\text{exc}}} \right)} \approx \sqrt{\Delta_0^2 + 4 \Omega_R^2} + \frac{\Delta_0}{\sqrt{\Delta_0^2 + 4 \Omega_R^2}} \frac{P^2}{2} \left( \frac{c}{n \hbar \pi L_c^{-1}} - \frac{1}{M_{\text{exc}}} \right)
\end{equation}

\medskip
\par
\noindent
The Taylor-expanded polariton energies are

\begin{equation}
\begin{aligned}
\varepsilon_{\text{LP/UP}}(P) \approx & \; \frac{1}{2} \left( \frac{c}{n} \hbar \pi L_c^{-1} + E_{\text{band}} - E_{\text{binding}} \right) 
\mp \frac{1}{2} \sqrt{\Delta_0^2 + 4 \Omega_R^2} \\
& + \frac{P^2}{4} \left[ \left( \frac{c}{n \hbar \pi L_c^{-1}} + \frac{1}{M_{\text{exc}}} \right) 
\mp \frac{\Delta_0}{\sqrt{\Delta_0^2 + 4 \Omega_R^2}} \left( \frac{c}{n \hbar \pi L_c^{-1}} - \frac{1}{M_{\text{exc}}} \right) \right]
\label{UPLPLin}
\end{aligned}
\end{equation}

\noindent
The dispersions of the photon, exciton, lower, and upper polaritons is visualised below.

\begin{figure}[H]
    \centering
    \includegraphics[width=0.55\linewidth]{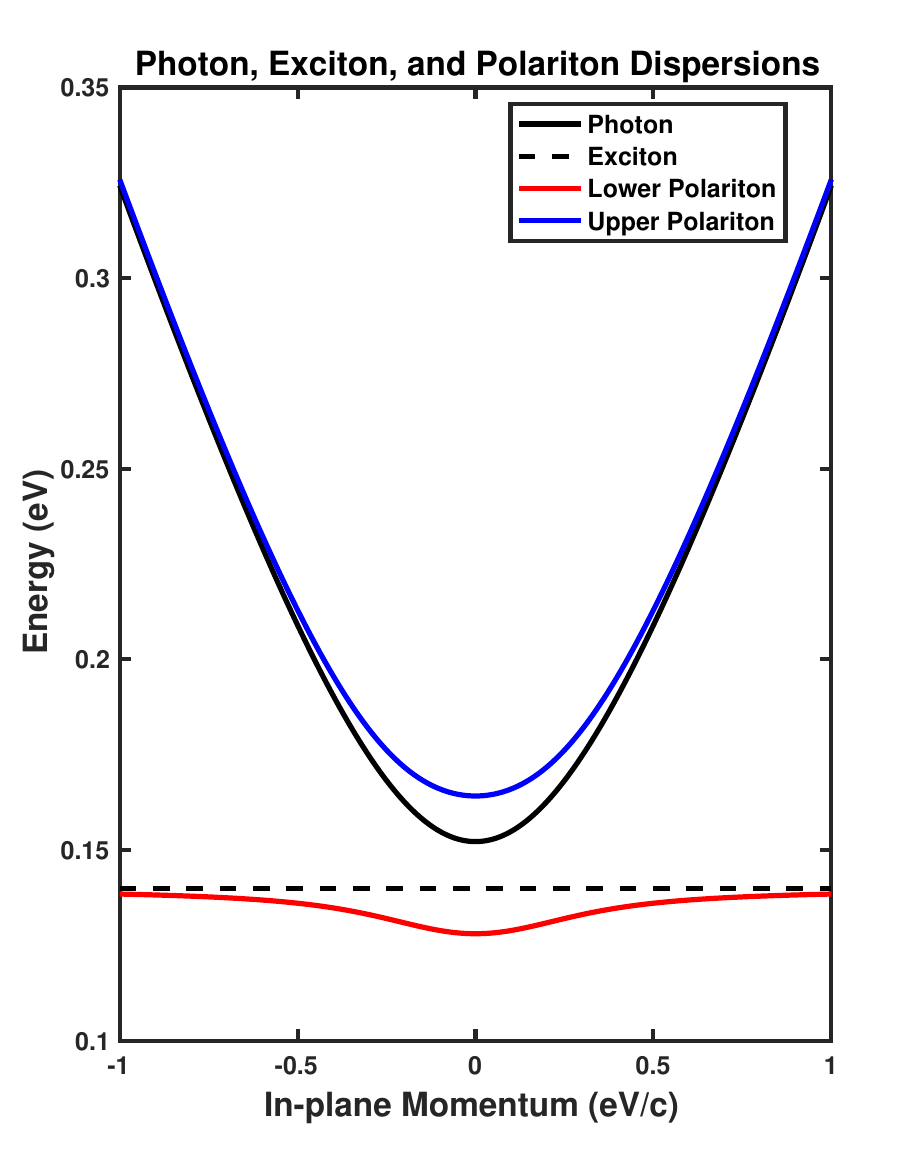}
    \caption{Energy dispersions of the photon, exciton, lower polaritons, and upper polaritons as a function of the in-plane momentum in one of our candidate materials - GaAs. The parameters used were $\Omega_R = 17 $ meV, a bare photon of energy 1.6~eV, and a finely adjusted detuning. The energy is measured relative to the top of the valence band.}
    \label{fig:dispersions}
\end{figure}

\medskip
\par
\noindent
The effective masses of the lower and upper polaritons can be found using the effective mass approximation, which relates the inverse effective mass to the second derivative of the energy spectrum with respect to momentum. The general formula is given by
\begin{equation}
\frac{1}{m_{\text{eff}}} = \frac{\partial^2 \varepsilon_{\text{LP/UP}}}{\partial P^2}
\end{equation}

\medskip
\par
\noindent
The Taylor-expanded expression for the polariton energy near small momentum \( P \) is given by Eq. (\ref{UPLPLin}). The term proportional to \( P^2 \) contributes to the effective mass calculation. We denote this coefficient by \( A_{\text{LP/UP}} \), where

\begin{equation}
A_{\text{LP/UP}} = \frac{1}{4} \left[ \left( \frac{c}{n \hbar \pi L_c^{-1}} + \frac{1}{M_{\text{exc}}} \right) 
\mp \frac{\Delta_0}{\sqrt{\Delta_0^2 + 4 \Omega_R^2}} \left( \frac{c}{n \hbar \pi L_c^{-1}} - \frac{1}{M_{\text{exc}}} \right) \right]
\end{equation}

\medskip
\par
\noindent
The effective mass is then 
\begin{equation}
\frac{1}{m_{\text{LP/UP}}} = 2 A_{\text{LP/UP}}
\end{equation}

\medskip
\par
\noindent
We find the effective masses for the lower and upper polaritons as
\begin{equation}
\frac{1}{m_{\text{LP}}} = \frac{1}{2} \left[ \left( \frac{c}{n \hbar \pi L_c^{-1}} + \frac{1}{M_{\text{exc}}} \right) 
- \frac{\Delta_0}{\sqrt{\Delta_0^2 + 4 \Omega_R^2}} \left( \frac{c}{n \hbar \pi L_c^{-1}} - \frac{1}{M_{\text{exc}}} \right) \right]
\label{LowerPolMass}
\end{equation}
\begin{equation}
\frac{1}{m_{\text{UP}}} = \frac{1}{2} \left[ \left( \frac{c}{n \hbar \pi L_c^{-1}} + \frac{1}{M_{\text{exc}}} \right) 
+ \frac{\Delta_0}{\sqrt{\Delta_0^2 + 4 \Omega_R^2}} \left( \frac{c}{n \hbar \pi L_c^{-1}} - \frac{1}{M_{\text{exc}}} \right) \right]
\label{UpperPolMass}
\end{equation}

\medskip
\par
\begin{figure}[H]
\centering \subfigure(a){
\includegraphics[width=0.45\textwidth]{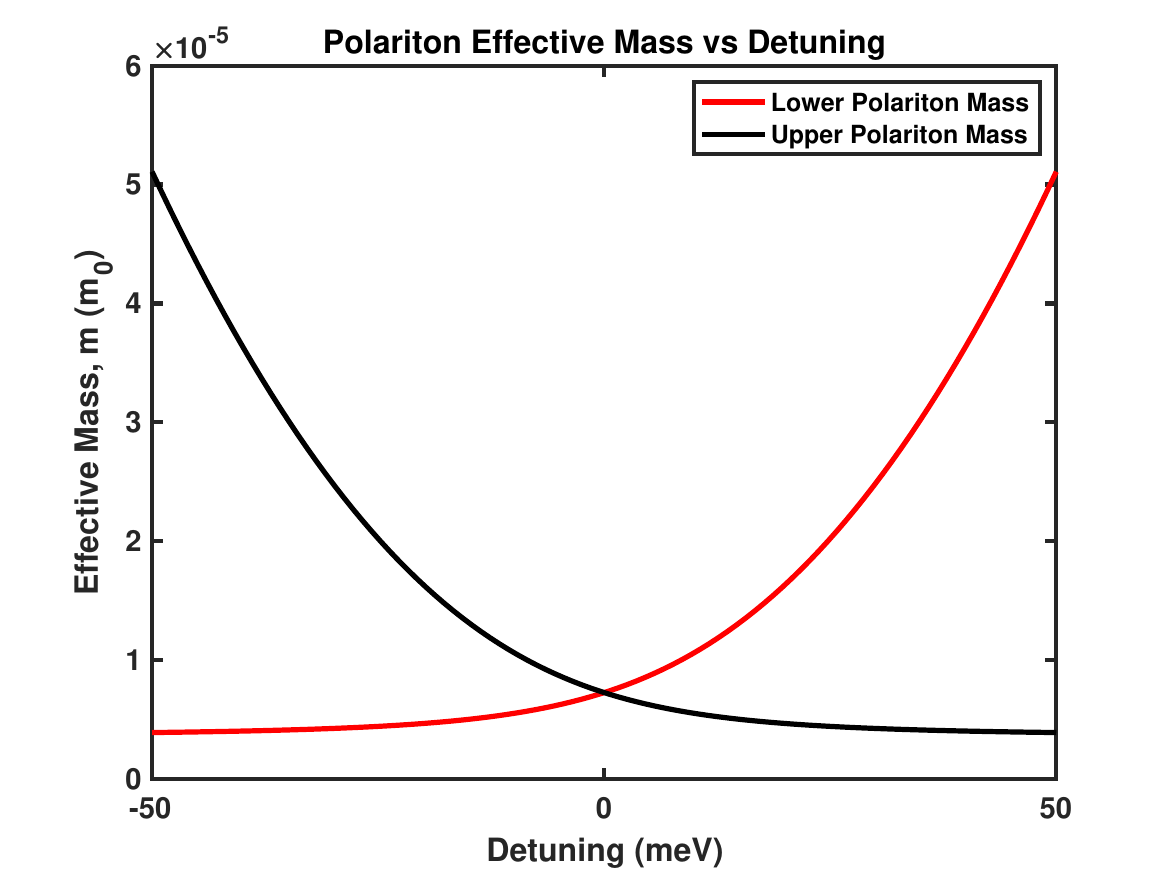}
} \subfigure(b){
\includegraphics[width=0.45\textwidth]{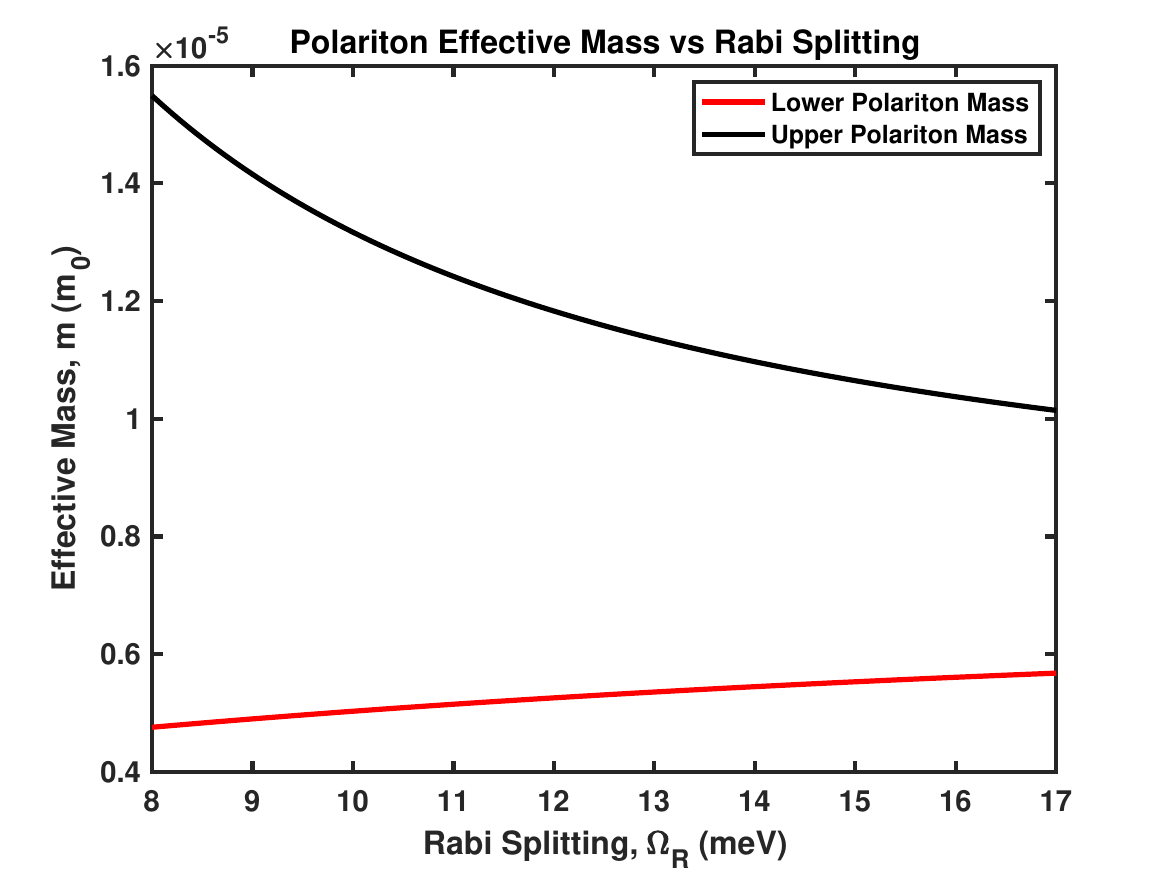}
}
\caption{(Color online) The polariton effective masses in GaAs as a function of (a) the detuning and (b) the Rabi splitting $\Omega_R$. For (a), we kept $\Omega_R = 15$ meV as a constant, and for (b) we set the detuning to be $\Delta_0$ = -10 meV.}
\label{MLUP}
\end{figure}

\noindent
We plot the effective masses of the upper and lower polaritons as a function of the detuning and the Rabi splitting. In Figure~\ref{MLUP} below we plot these dependances.

\medskip
\par
\noindent

It is worth noting that the effective masses of the upper and lower polariton are constant and identical at a detuning of 0. Away from the detuning of 0, the Rabi splitting allows for the masses of the upper and lower polaritons can be adjusted as a function of the Rabi splitting.

\section{Results for GaAs/AlGaAs quantum wells in a GaAs microcavity}
\label{resultsGaAs}
  
\begin{figure}[H]
    \centering
    \subfigure[\;]{
        \includegraphics[width=0.4\textwidth]{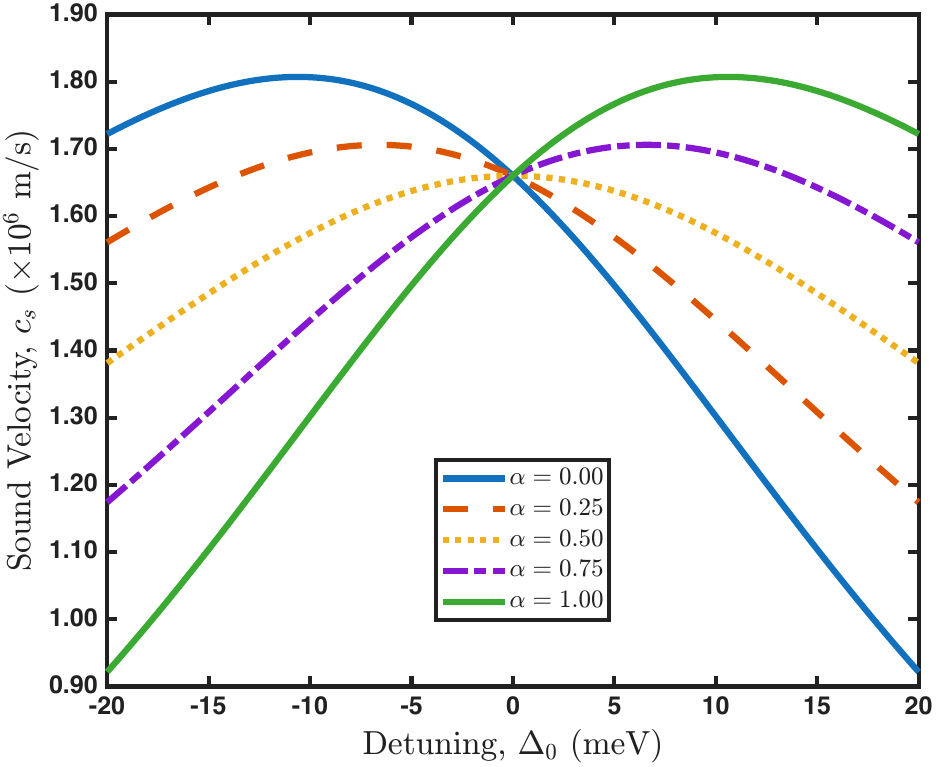}
        \label{fig:SoundVelocityVersusDetuning}
    }
    \hfill
    \subfigure[\;]{
        \includegraphics[width=0.4\textwidth]{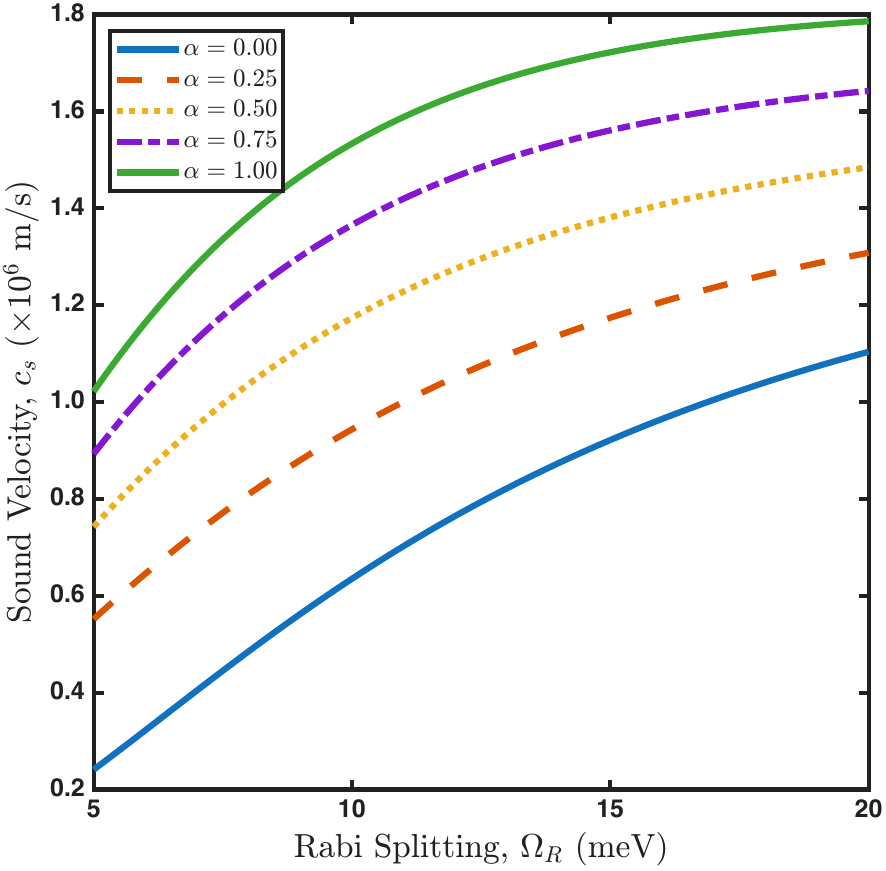}
        \label{fig:SoundVelocityVersusRabi}
    }
    \caption{Sound velocity $c_s$ of a two-component polariton condensate:
    (a) dependence on detuning $\Delta_0$ at fixed $\Omega_R = 15$~meV, 
    and (b) dependence on Rabi splitting $\Omega_R$ at fixed $\Delta_0 = 20$~meV. 
    The total polariton density is $n_0 = 1\times10^{11}~\text{cm}^{-2}$.}
    \label{fig:SoundVelocityDetuningAndRabiGaAs}
\end{figure}

We plot in this section all analogous results for GaAs/AlGaAs quantum wells in a GaAs microcavity. We begin by presenting plots of the sound velocity as a function of detuning and the Rabi splitting in Figure~\ref{fig:SoundVelocityDetuningAndRabiGaAs}.

Similarly we plot the sound velocity as a function of the total polariton density in Figure~\ref{fig:SoundVelocityConcAndAlphaGaAs} below.

\begin{figure}[H]
    \centering
    \includegraphics[width=0.45\textwidth]{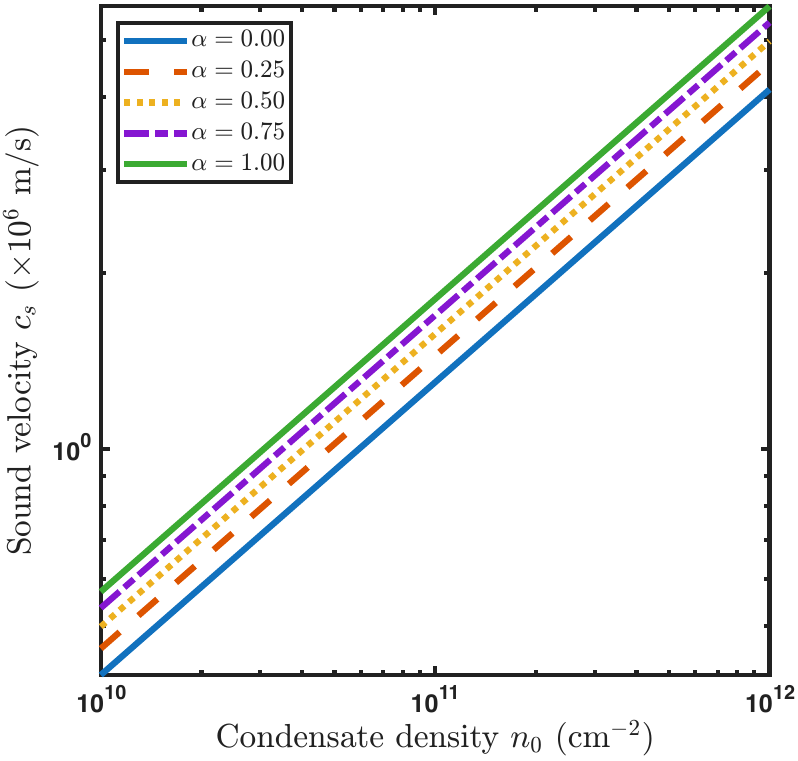}
    \caption{Sound velocity $c_s$ as a function of total polariton density $n_0$ for different population-split parameters $\alpha$. 
    The Rabi splitting and detuning are fixed at $\Omega_R = 15$~meV and $\Delta_0 = 10$~meV, respectively.}
    \label{fig:SoundVelocityConcAndAlphaGaAs}
\end{figure}

Next we present the results of the critical temperature $T_c$ as a function of the detuning and Rabi splitting, in Figure~(\ref{fig:CriticalTempDetuningAndRabiGaAs}) below.

\begin{figure}[H]
    \centering
    \subfigure[]{
        \includegraphics[width=0.45\textwidth]{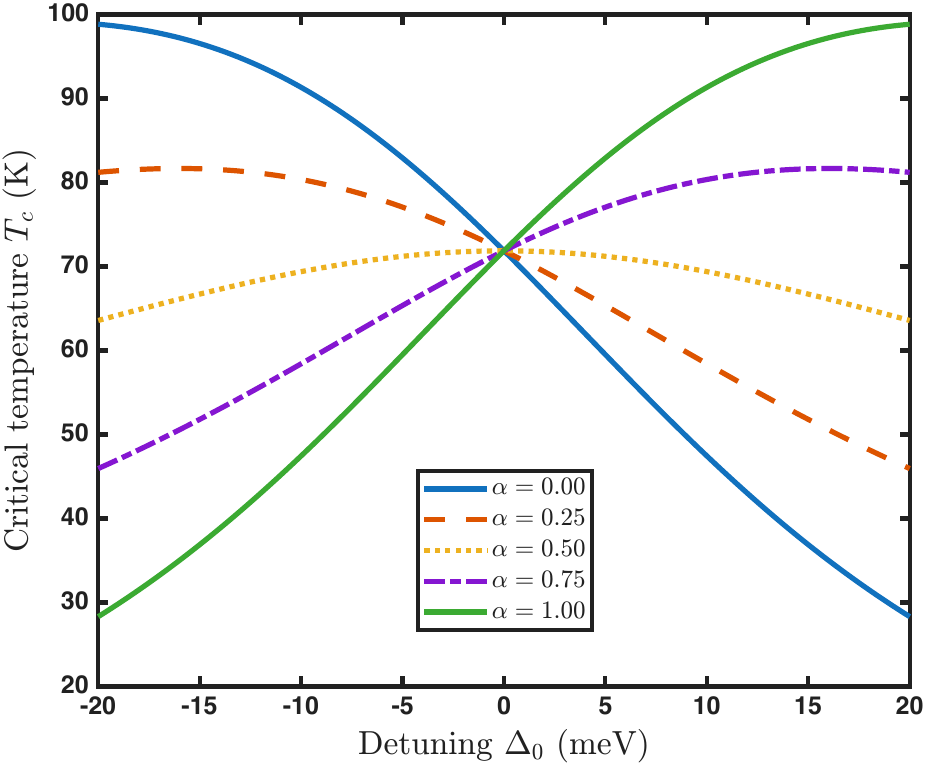}
        \label{fig:TempVsDetuning}
    }
    \hfill
    \subfigure[]{
        \includegraphics[width=0.45\textwidth]{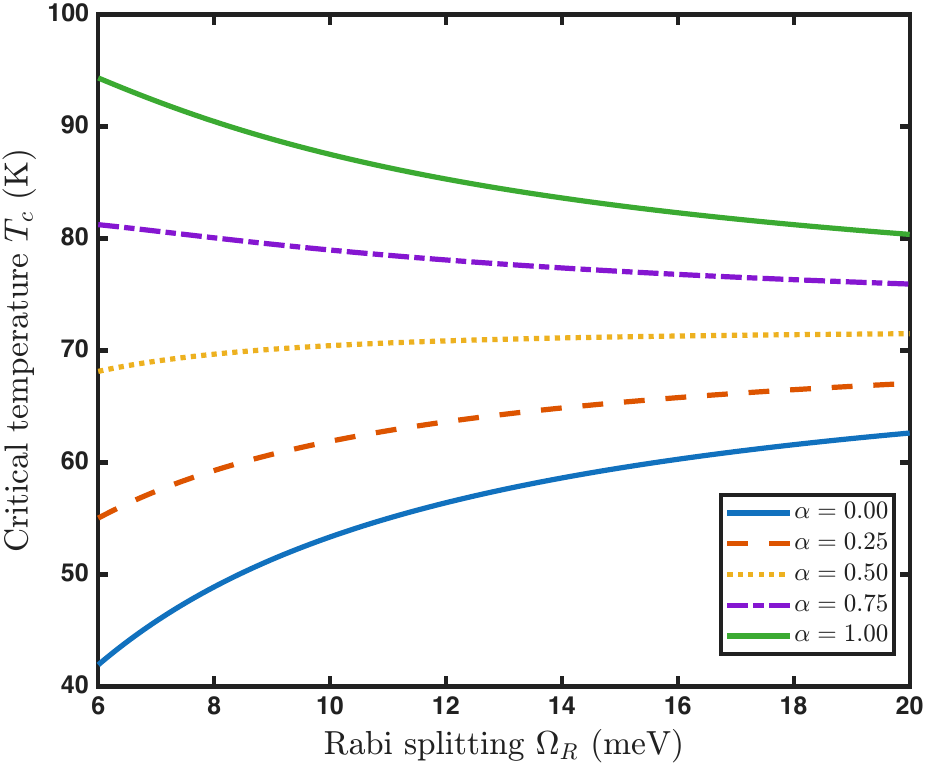}
        \label{fig:TempVsRabi}
    }
    \caption{Critical temperature $T_c$ in a two-component polariton condensate. 
    (a) Dependence on detuning $\Delta_0$ at fixed Rabi splitting $\Omega_R = 15$ meV. 
    (b) Dependence on Rabi splitting $\Omega_R$ at fixed detuning $\Delta_0 = 5$ meV. 
    In both panels, the total polariton density is held constant at  $n_0 = 1 \times 10^{11}~\text{cm}^{-2}$, and results are shown for multiple values 
    of the population--split parameter $\alpha$.}
    \label{fig:CriticalTempDetuningAndRabiGaAs}
\end{figure}

Afte that we present the results of the critical temperature $T_c$ as a function of the total polariton density in Figure~(\ref{fig:CriticalTempConcGaAs}) below.

\begin{figure}[H]
    \centering
    \subfigure[]{
        \includegraphics[width=0.45\textwidth]{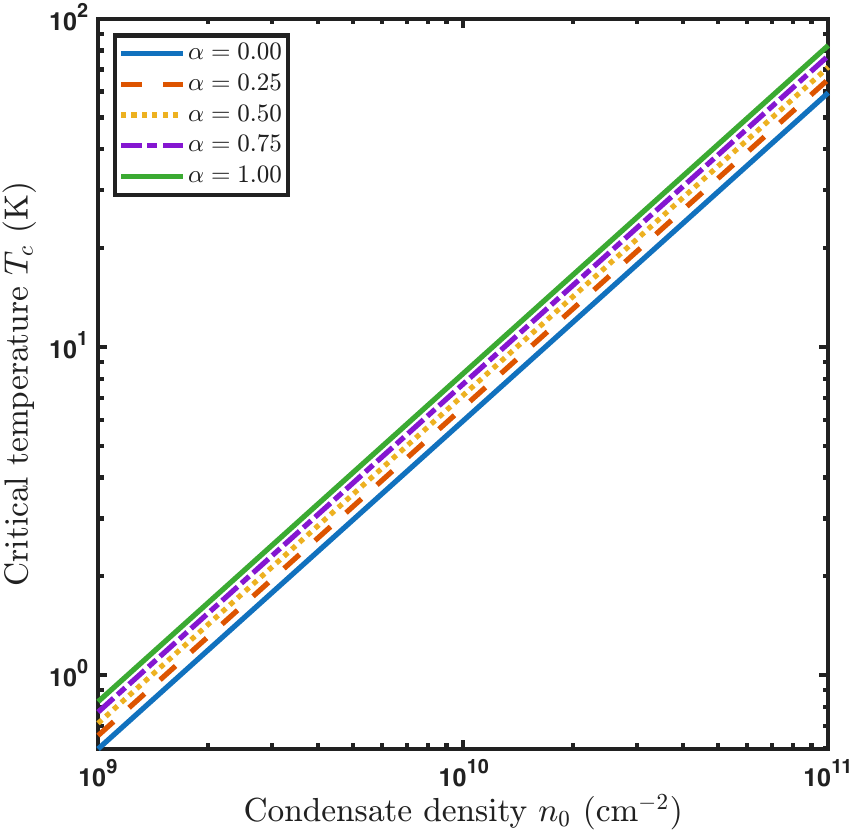}
        \label{fig:TempVsConc}
    }
    \caption{Critical temperature $T_c$ in a two-component polariton condensate as a function of the total polariton density $n_0$ for multiple values of the condensate population--split parameter $\alpha$. The Rabi splitting is fixed at $\Omega_R = 15$ meV and the detuning at $\Delta_0 = 5$ meV.}
    \label{fig:CriticalTempConcGaAs}
\end{figure}

Finally we present the phenomenological inference plots for GaAs/AlGaAs below in Figure~\ref{fig:AlphaVsNcAndTcGaAs}.
 
\begin{figure}[H]
    \centering
    \subfigure[]{
        \includegraphics[width=0.45\textwidth]{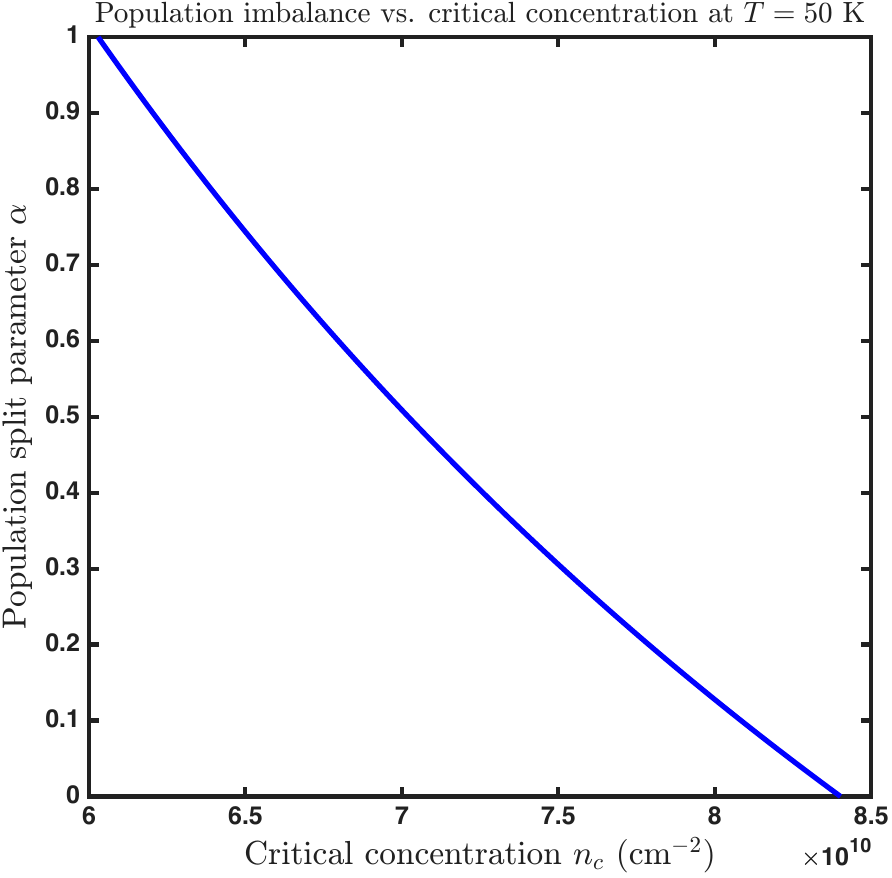}
        \label{fig:AlphaVsNc}
    }
    \hfill
    \subfigure[]{
        \includegraphics[width=0.45\textwidth]{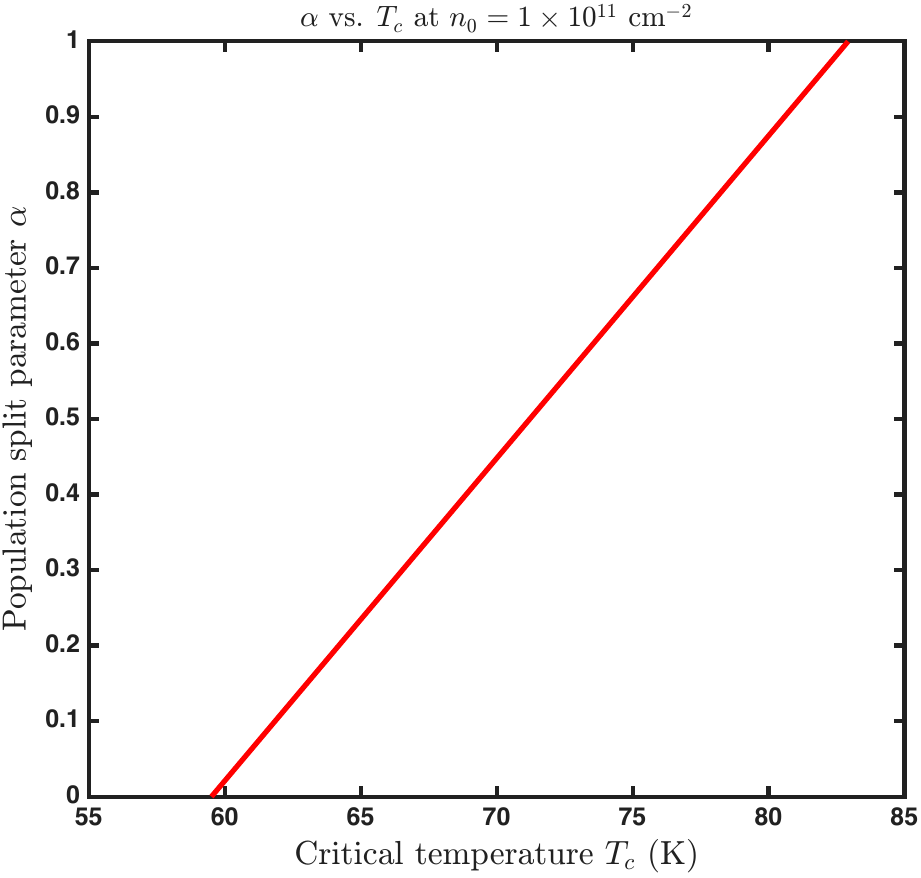}
        \label{fig:AlphaVsTc}
    }
    \caption{Condensate population--split parameter $\alpha$ in a two-component polariton condensate. 
    (a) Dependence of $\alpha$ on the critical concentration $n_c$ at fixed system temperature 
        $T=50$~K, Rabi splitting $\Omega_R = 15$~meV, and detuning $\Delta_0 = 5$~meV. 
    (b) Dependence of $\alpha$ on the critical temperature $T_c$ at fixed total polariton density 
    $n_0 = 1 \times 10^{11}~\text{cm}^{-2}$, Rabi splitting $\Omega_R = 8$~meV, and detuning 
    $\Delta_0 = 10$~meV.}
    \label{fig:AlphaVsNcAndTcGaAs}
\end{figure}

\end{document}